\documentclass[3p, review]{elsarticle}

\usepackage{ecrc}

\volume{00}

\firstpage{1}

\journalname{New Astronomy}

\runauth{Giorgi et al.}

\jid{NewA}

\jnltitlelogo{Procedia Computer Science}

\CopyrightLine{2011}{Published by Elsevier Ltd.}

\usepackage{amssymb}
\usepackage[figuresright]{rotating}

\usepackage[utf8x]{inputenc}
\usepackage{multirow}
\usepackage{graphicx}
\usepackage[section] {placeins} % Este paquete hace que las figuras no se muevan tanto
\begin{document}

\begin{frontmatter}



\dochead{}
%% e.g. \dochead{17th International Conference on Dynamical Processes in Excited States of Solids}

\title{The nearby Galaxy structure toward the Vela Gum nebula}


\author[label1]{Giorgi, E. E.\corref{cor1}}
\ead{egiorgi@fcaglp.unlp.edu.ar}
\author[label1]{Solivella G. R.\corref{cor1}}
\ead{gladys@fcaglp.unlp.edu.ar}
\author[label1,label2]{Perren G. I.}
\ead{gperren@fcaglp.unlp.edu.ar}
\author[label1]{V\'azquez R. A.}
\ead{rvazquez@fcaglp.unlp.edu.ar}

\cortext[cor1]{Visiting Astronomer, Complejo Astron\'omico El Leoncito
operated under agreement between the Consejo Nacional de
Investigaciones Cient\'ificas y T\'ecnicas de la Rep\'ublica Argentina and
the National Universities of La Plata, C\'ordoba and San Juan.}

\address[label1]{Facultad de Ciencias Astron\'omicas y Geof\'isicas, UNLP,
IALP-CONICET, La Plata, Argentina}
\address[label2]{Instituto de F\'isica de Rosario, IFIR (CONICET-UNR),
Parque Urquiza, 2000 Rosario, Argentina}

\begin{abstract}
We report on $UBVI$ photometry and spectroscopy for MK
classification purposes carried out in the fields of five open
clusters projected against the Vela Gum in the Third Galactic
Quadrant of the Galaxy. They are Ruprecht 20, Ruprecht 47,
Ruprecht 60, NGC 2660 and NGC 2910. We could
improve/confirm the parameters of these objects derived before.
Ruprecht 20 is not a real physical entity, in agreement with
earlier suggestions. Ruprecht 47, a young cluster in the Galactic
plane, at 4.4 kpc from the Sun is quite farther than in previous
distance estimations and becomes, therefore, a member of the
Puppis OB2 association. For the first time Ruprecht 60 was
surveyed in $UBVI$ photometry. We found it to be placed at 4.2 kpc
from the Sun of about and 1 Gyr old. NGC 2660 is another old
object in our survey for which distance and age are coincident
with previous findings. NGC 2910 turns out to be a young
cluster of Vela OB1 association at a distance of 1.4 kpc
approximately and 60 Myr old. The spectroscopic parallax method
has been applied to several stars located in the fields of four
out of the five clusters to get their distances and reddenings.
With this method we found two blue stars in the field of NGC 2910
at distances that make them likely members of Vela OB1 too. Also,
projected against the fields of Ruprecht 20 and Ruprecht 47 we
have detected other young stars favoring not only the existence of
Puppis OB1 and OB2 but conforming a young stellar group
at $\sim1$ kpc from the Sun and extending for more than 6 kpc
outward the Galaxy. If this is the case, there is a thickening of
the thin Galactic disk of more than 300 pc at just 2-3 kpc from
the Sun. Ruprecht 60 and NGC 2660 are too old objects that have no
physical relation with the associations under discussion. An
astonishing result has been the detection in the background of
Ruprecht 47 of a young star at the impressive distance of 9.5 kpc
from the Sun that could be a member of the innermost part of the
Outer Arm. Another far young star in the field of NGC 2660, at
near 6.0 kpc, may become a probable member of the Perseus Arm or
of the inner part of the Local Arm. The distribution of young
clusters and stars onto the Third Galactic Quadrant agrees with
recent findings concerning the extension of the Local Arm as
revealed by parallaxes of regions of star formation. We show
evidences too that added to previous ones found by our group
explain the thickening of the thin disk as a combination of flare
and warp.
\end{abstract}


\end{frontmatter}

% \linenumbers

\section{Introduction}

This investigation forms part of a long term project aimed
at understanding the nearby spiral structure in the Third Galactic
Quadrant of the Milky Way (hereinafter TGQ) from an optical point
of view by means of open clusters. The still unresolved question
about the grand design structure of the Milky Way (see, e.g. Choi
et al. 2014) has been demanding an extraordinary observational
effort during the last decades. Actually, a crucial starting point
took place in the $^\prime70$ when Georgelin \& Georgelin (1976)
produced the first map of the Galaxy utilizing HII regions as
spiral tracers. This first map showed that a four arms model fits
the observations very well, a concept widely used yet. Since then,
a number of other maps showing the structure of the Galaxy have
been developed by means of different spiral tracers such as
molecular clouds (Efremov 1998) or by combining them with HII
regions (Hou et al. 2009). Dame et al. (2001) collected CO
observations from several sources to map the entire Milky Way.
While Russeil (2003) carried out a multiwavelength study of the
plane of our Galaxy and found evidences that favored a four arms
model, Nakanishi \& Sofue (2003) presented a three dimensional map
constructed with HI observations concluding that a three arms
model is better to fit the observations. Churchwell et al. (2009)
combined data from a variety of sources in the wavelength range
from 3.6 to 24 microns adopting also a four arms model but
recognizing some difficulties of interpretation. This is, since
the controversy about the number of spiral arms in the Galaxy is
still open we suggest the readers to have a look to the summary of
spiral arms parameters in the Vall\'ee (2013, 2014) paper series
and references therein. However, in almost all these cases distances
depend on a rotational curve and the adopted kinematic model.
Thus, strong ambiguities take place and only can be resolved by
adopting other tracers of spiral structures with distances
independent from the properties of the models. In this respect,
trigonometric parallaxes and proper motions of massive star
forming regions in combination with information from HII regions,
giant molecular cloud and masers is a promising initiative since
distance ambiguities can be removed (Reid et al. 2009, 2014; Hou
\& Han (2014)). Just to mention, methanol and water masers have
been successfully used for some authors to study particular
regions of the Milky Way such as the Local, Sagittarius and
Perseus Arms (Xu et al. (2014), Wu et al. 2014, Choi et al. 2014,
among others).

In this framework, young open clusters become excellent
spiral tracers since they can be observed at large distances. The
additional advantage of using open clusters is that their
distances do not rely on any rotational curve and kinematic model.
Contrarily, observations and detection of open clusters are highly
dependent of the visual extinction along the line of view in the
Galactic plane. Several investigations biased to study the large
structure of the Milky Way using open clusters and stellar
component were developed by Moffat \& Vogt (1973) and Vogt \&
Moffat (1975) although with a strong magnitude cut-off. This
precluded from arriving to conclusive definitions about the
extension of the Local Arm (Orion), the angular
recognition (and the real entity) of the Perseus Arm and the
possible trace of the Outer Arm (Norma-Cygnus). Some other efforts
were carried out by McCarthy \& Miller (1974), Moffat \&
FitzGerald (1974), Clari\'a (1974), Reed \& FitzGerald (1984ab),
Slawson \& Reed (1988), Reed \& FitzGerald (1985), Reed (1990),
all of them concerning the distribution and structure shown by the
stellar components and also by open clusters. Most of the optical
inspections performed in the TGQ were concentrated along the
region $240^\circ$ to $280^\circ$ in longitude, including the
Puppis and Vela regions. This could be done, in part, thanks to
the presence of a dust-window, the so-called Fitzgerald window
(FitzGerald 1968) near the Puppis association. A thorough
information on related earlier works can be found in Reed (1989)
but we also suggest the contribution from Kaltcheva \& Hilditch
(2000) who revealed the presence of very distant blue stars in the
longitude interval from $215^\circ < l < 275^\circ$. Indeed, the
absorption has been the greatest obstacle to observations. During
the last two decades a notorious feature has been confirmed in the
TGQ from the dust mapping for Schlegel et al. (1998) together with
the Hydrogen emission maps from SHASSA; these two surveys suggest
a symmetry brake-up in the sense that both, dust and emission
(neutral gas) show a tendency to locate below the formal Galactic
plane at $b = 0^\circ$ Galactic latitude. This fact is in
agreement with previous investigations where it has been shown
that the HI layer in the Galaxy flares, becomes thicker at larger
distances from the Galactic Center (Henderson et al. 1982;
Kulkarni et al. 1982; Burton \& te Lintel Hekkert 1986) on a side.
On the other, a tendency for the TGQ gas clouds to fall below the
Galactic plane has been revealed and interpreted as evidence for
warp in the molecular disk as well (May, Murphy \& Thaddeus 1988;
May et al. 1997). CO observations of the Infrared Astronomical System
(IRAS) sources Wouterloot et al. (1990) concluded that, although
the HI layer is substantially thicker than the molecular clouds
layer in the inner Galaxy, the z-height distribution of the
molecular clouds show the same thickness in the outer Galaxy. A
few years ago our group (V\'azquez et al. 2008, Carraro et al.
2005, Moitinho et al. 2006, Carraro et al. 2015) presented and
discussed a new picture of the spiral structure in the TGQ built
by means of a mixture of optical data of open clusters, field blue
stars and CO sources distributed along 60 degrees in Galactic
longitude. This new picture confirms the presence of the warp (in
optical and CO), reveals the existence of the Outer Arm and
suggests that the Local Arm extends from the Sun to its encounter
with the Outer arm. However, none of these studies can give
definite evidences about the trace of the Perseus Arm except for a
very few open clusters that, ambiguously, can be related to the
Local Arm too.

Our ongoing observational programme at La Plata
Observatory was designed to improve parameters of known clusters
and to look for those with no information at all. This is an
essential task that allows us to find far spiral tracers and
recognize the grand design spiral structure in the TGQ. So, we
report here on results coming from CCD $UBVI$ photometry combined
with spectroscopic measures for five open clusters, Ruprecht 20,
Ruprecht 47, Ruprecht 60, NGC 2660 and NGC 2910, all of them
projected against the Vela Gum Nebula at $240^\circ < l < 260^\circ$
including Vela and Puppis associations. To situate the readers the
Gum Nebula, one of the biggest H\.II regions of our Galaxy, covers
an angular extension of $\sim 36^\circ$ and is centered at
$l^\circ=258$, $b^\circ=-5$. Sahu (1992) analyzed the interstellar
medium in the zone up to 2 kpc in Vela-Puppis; cometary globules
and the star formation process in the Gum Nebula have also been
studied by Kim et al. (2005). Definitely, $\zeta$ Pup (O4If) and
$\gamma^2$ Vel (WC8+O7.5 I), the OB associations, Trumpler 10 and
Vela OB2, seem to be the responsible entities of most of the
photo-ionization in the nebula itself. To be remarked is the
presence of the IRAS Vela Shell (Sahu 1992), an expanding dust
shell surrounding Vela OB2 with 6-8 pc radius enclosing the Gum
Nebula. Behind it, the Vela Molecular Ridge (Murphy \& May 1991),
a superposition of giant CO molecular clouds, appears obscuring
the farthest part of the Galaxy.

The purpose of the present investigation is twofold: on a
side we want to re-discuss the parameters of five clusters in the
TGQ onto the basis of a new set of photometry (one of them has
never before been subject of $UBVI$ photometry). On the other
side, we want to connect the properties of these objects with the
structure of Puppis and Vela associations and look for evidences
of the grand design spiral structure. The photometric observations
have been complemented with spectroscopy of several stars located
in the cluster regions to discard field interlopers but also to
estimate the absorption law. Table 1 lists coordinates of the
clusters together with their angular sizes (from our analysis, in
advance).

The remaining of the paper is organized as follows. In \S~2 we
present and discuss the data acquisition of photometry and
spectroscopy for five open clusters projected against the Vela Gum
nebula. In \S~3 we illustrate the procedure applied for the
derivation of fundamental parameters for the cluster sample. \S~4
presents a cluster- by- cluster discussion and the trend
color excess. Discussion of the derived Galactic structure and
our conclusions are given in \S~5 and 6, respectively.

\section{Observations and data reduction}

\subsection{Photometric data.}

\begin{center}
\begin{figure}[ht!]
\begin{flushleft}
% \fbox{}
\resizebox{16cm}{!}{\includegraphics{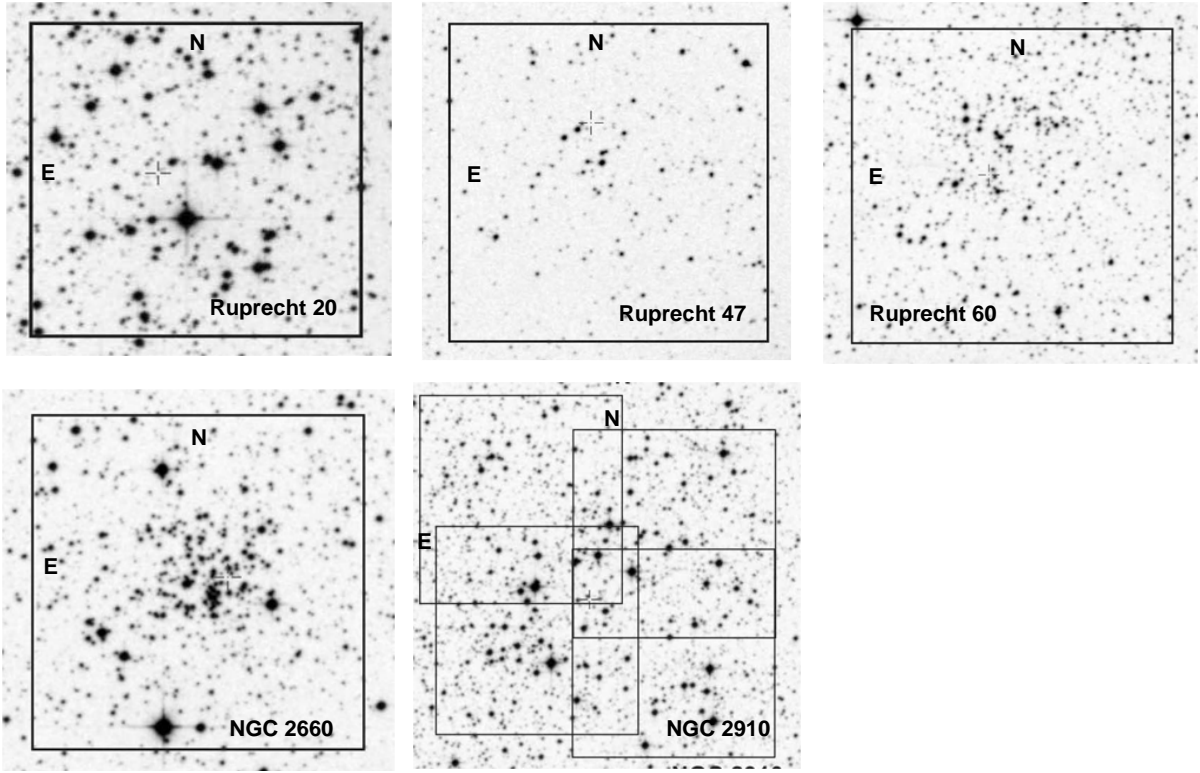}}
\caption{\small{Observed areas of the five clusters projected
against {\it Digital Sky Survey} (DSS) images. Photometry has
been performed within each
black square $\sim 4^\prime$ on a side. North and East are
indicated.}} \label{cumulos_perfiles}
\end{flushleft}
\end{figure}
\end{center}

We show in Fig. 1 the surveyed areas with CCD $UBV(I)$ photometry
of the five clusters in our sample enclosed by squares $\sim
4^\prime$ on a side. NGC 2660 and NGC 2910 were measured at Las
Campanas Observatory, Chile, on February 1996 with the 0.6-m
telescope of the University of Toronto Southern Observatory (UTSO)
equipped with a CCD PM $512\times512$ METHACROME-II UV coated
camera. In these two clusters we carried out $UBVI$ photometry.
The scale of the CCD camera is $0.45^{\prime\prime}/pix$ what
generates a field of $4^{\prime}$ on a side. The nights were
photometric with seeing values between $1.2^{\prime\prime}$ and
$1.4^{\prime\prime}$. Four frames were necessary in the case of
NGC 2910 to get the best possible coverage of the cluster.

\begin{table}[htb!]
\begin{center}
\caption{Cluster's radii from Subsection 3.1.}
 \begin{tabular}{lllllc}
\hline
\hline
Cluster & $\alpha_{2000}$ & $\delta_{2000}$ &$l^{\circ}$&$b^{\circ}$& radius [$^\prime$] \\
\hline
Ruprecht 20 &  07 26 43 & $-$28 49 00& 242.451 & $-$5.752 & - \\
Ruprecht 47 &  08 02 19 & $-$31 04 00& 248.252 & $-$0.188 & 4 \\
Ruprecht 60 &  08 24 27 & $-$47 13 00& 264.106 & $-$5.509 & 4 \\
NGC 2660    &  08 42 38 & $-$47 12 00& 265.929 & $-$3.013 & 6 \\
NGC 2910    &  09 30 30 & $-$52 55 06& 275.310 & $-$1.176 & 6 \\
\hline
\end{tabular}
\label{radius}
\end{center}
\end{table}

CCD $UBV$ photometry in Ruprecht 20, Ruprecht 47 and CCD $UBVI$
observations in Ruprecht 60 were carried out at the CASLEO
Observatory (Argentina) using the 2.15-m telescope equipped with a
CCD ROPER 1300B, $1340\times1300$ pixels and
$0.226^{\prime\prime}/pix$, covering $4.2^{\prime}$ on a side.
Ruprecht 47 was observed on the night March 9 (2005), Ruprecht 60
on April 3 (2005) and Ruprecht 20 on January 3 (2006). The typical
seeing values for these observing runs varied from $1.3~''$
to $2.1~''$.

Details of exposure times and number of observations per filter
for the five regions observed can be found in Table 2. In all
cases, very short exposures were necessary to avoid
saturation of the brightest stars.

\begin{table}[htb!]
\begin{center}
\caption{Exposure times for all clusters in seconds. In
parenthesis is the number of averaged images.}
 \begin{tabular}{llrrrr}
\hline
\hline
Cluster & Exp time & \multicolumn{4}{c}{Passbands} \\
 \hline
& & $U$ & $B$ & $V$ & $I$ \\
\hline

\multirow{2}{*}{Ruprecht 20} & $medium$ & & 3 & 3 & -\\
& $long$ & 300 & 300 & 300 & -\\
\hline

\multirow{3}{*}{Ruprecht 47} & $short$  &  &   & 10 & 2\\
& $medium$ & & 60 & 60 & 20\\
& $long$ & 300 & 300 & 300 & 130\\
\hline

\multirow{2}{*}{Ruprecht 60} & $medium$ & & 30 & 30 & 20\\
& $long$ & 300 & 300 & 300 & 130\\
\hline

\multirow{3}{*}{NGC 2660} & $short$  &  & 200  & 100  & 2  \\
& $medium$ & &  &  & 20 \\
& $long$ &1000 (2)& 1000 & 600 & 130 \\
\hline

\multirow{3}{*}{NGC 2910} & $short$  &200  & 30  & 10  & 2  \\
& $medium$ & & 200 & 100 & 20 \\
& $long$ &1000 (2)& 1000 & 600 & 130 \\
\hline
\end{tabular}
\label{exp_times}
\end{center}
\end{table}

Instrumental magnitudes were obtained by adjusting the PSF using
the DAOPHOT package (Stetson 1987) included within IRAF. To carry
the instrumental magnitudes to the standard $UBVI$ system, the
following fields of standard stars were used:

\begin{itemize}
 \item Ruprecht 20, standard stars from the list of Landolt (1992)
 and stars observed by Vogt and Moffat (1972) as secondary standards.
 \item Ruprecht 47, standard stars from the list of Landolt (1992), using the 10 stars observed by
Vogt and Moffat (1972) in this cluster as secondary calibrators.
 \item Ruprecht 60, standard stars from the list of Landolt (1992).
 \item NGC 2910 and NGC 2660, a set of over 30 stars observed in open clusters NGC 5606
(V\'azquez et al. 1994) and Trumpler 18 (V\'azquez \& Feinstein
1990) was used.
\end{itemize}

The transformation equations to the standard system were of the
form:

\begin{center}
\noindent
\hspace*{0.5cm} $u=U+u_1+u_2 \times X+u_3 \times (U-B)$ \\
\hspace*{0.5cm} $b=B+b_1+b_2 \times X+b_3 \times (B-V)$ \\
\hspace*{0.5cm} $v=V+v_1+v_2 \times X+v_3 \times (B-V)$ \\
\hspace*{0.5cm} $i=I+i_1+i_2 \times X+i_3 \times (V-I)$ \smallskip
\end{center}

\noindent where $u_2$, $b_2$, $v_2$ and $i_2$ are the extinction
coefficients for the $UBVI$ bands, $X$ are the air masses for each
exposure and $u_1$, $b_1$, $v_1$, $i_1$, $u_3$, $b_3$, $v_3$, and
$i_3$ the fitted parameters. The extinction coefficients were
taken from Grotues \& Gocherman (1992) for observations at Las
Campanas Observatory and from Giorgi et al. (2007) for the CASLEO
observations. Final photometric tables are available at CDS.

\begin{table}[htb!]
\caption{Differences with previous works in the sense this work
minus other authors. N is the number of stars in common.}
\begin{tabular}{l|ccrrrr}
\hline \hline \multirow{2}{*}{Cluster} & \multirow{2}{*}{Author} &
\multirow{2}{*}{Data type} & \multirow{2}{*}{$\Delta V$} &
\multirow{2}{*}{$\Delta (B-V)$} &
\multirow{2}{*}{$\Delta (U-B)$} & \multirow{2}{*}{N} \\
&&&&&& \\
\hline

Ruprecht 20 & (1) & Photoelectric & $0.01\pm0.05$ & $0.00\pm0.03$ & $0.01\pm0.11$ & 11 \\
\hline

Ruprecht 47 & (1) & Photoelectric & $0.03\pm0.04$ & $0.04\pm0.01$ & $-0.03\pm0.03$ & 10 \\
\hline

\multirow{2}{*}{NGC 2660} & \multirow{2}{*}{(2)} & CCD ($V< 16m$) & $0.14\pm0.04$ & $-0.02\pm0.04$ & $-0.15\pm0.08$ & 163 \\
& & CCD ($V< 19m$) & $0.14\pm0.07$ & $0.00\pm0.10$ & $-0.20\pm0.10$ & 381 \\
\hline

\multirow{3}{*}{NGC 2910} & (3) & Photoelectric & $0.00\pm0.30$ & $-0.20\pm0.30$ & $0.20\pm0.60$ & 39 \\
& (4) & Photographic & $0.00\pm0.10$ & $0.00\pm0.10$ & $0.20\pm0.10$ & 37 \\
& (5) & CCD & $-0.04\pm0.09$ & $0.1\pm0.20$ & $0.12\pm0.07$ & 58$^{*}$ \\
\hline
\end{tabular}
\newline
References: \small{(1) Vogt and Moffat (1972); (2) Sandrelli et
al. (1999); (3) Becker (1960); (4) Topaktas (1981); (5) Ramsay \&
Pollaco (1992); $^\ast$ Ramsay \& Pollaco (1994) give $(U-B)$ data
for only 17 stars.}
\end{table}

Table 3 shows in self explanatory format differences in magnitude
and color indices together with the standard deviation of the mean
between photoelectric/photographic and CCD data from several
authors and our own data, except for Ruprecht 60, in the sense our
measures minus other authors. The number of stars used in each
comparison is indicated. Inspecting the differences we find an
excellent agreement of our photometry with photoelectric measures
for Ruprecht 20 and Ruprecht 47. However, strong magnitude and
color off-sets appear in the case of NGC 2660. In this respect we
want to mention that Sandrelli et al. (1999) do not mention they
compared their photometry of NGC 2660 with previous values except
for $V$, $B$ and $U$ isolated bands in a few stars (Sandrelli et
al. is the most extended photometric study to date). What we see
in Fig. 3 in Sandrelli et al. (1999) is a strong scatter around
mean values when they compare to Hartwick \& Hesser (1973) data.
We are confident with the procedures we routinely undertake in
doing photometry so we believe that the strong off-sets are due to
zero point problems in Sandrelli et al. (1999) photometry. As for
NGC 2910 we find a large discrepancy with Becker (1960) data and
also with CCD data from Ramsay \& Pollaco (1992). In the case of
Ruprecht 60 no comparison is available since this cluster has only
one study (Bonatto \& Bica, 2010) made with {\it J}, {\it H}, and
{\it K$_s$} bands of the Two-Micron All-Sky Survey (2MASS).

\subsection{Spectroscopic data and MK classification}

Spectra for MK classification purposes were collected at the
2.15-m telescope of CASLEO (Argentina) during several observing
runs. The purpose of obtaining spectra of field stars in
clusters is supported by the need of getting useful additional
information from them to estimate precisely the visual absorption
in each direction. These stars were selected primarily because of
their brightness and then for the degree of proximity to the
cluster center. They were observed according to the following
detail:

\begin{itemize}
 \item Ruprecht 20, 12 stars during three observing runs in January 2006, February 2007 and February 2008
 \item Ruprecht 47, 8 stars during two observing runs in January 2010 and February 2011
 \item NGC 2660, 15 stars throughout several observing runs between 2002 and 2006
 \item NGC 2910, 6 stars during two observing runs between 12-15 March (2002) and 12-13 March (2004)
\end{itemize}

Figure 2 shows the spectroscopic finding charts. Numbers in the
charts give the star identification used in this article. All
spectra were obtained with the REOSC Cassegrain spectrograph
attached to the 2.15-m telescope and the Tek $1024 \times 1024$
detector using a 600 l/mm grating in the first order. Spectra have
a wavelength coverage from 3900 {\AA} to 5500 {\AA} (the
traditional MK spectral region for classification from CaII K to H
lines), 2.5 {\AA}"/pixel dispersion (1800 resolution) and were reduced
using standard procedures with IRAF. Our strategy for reducing
cosmic rays and improve the signal/noise ratio consists in taking,
at least, two spectra for each star with exposure times between 20
and 40 minutes depending on the star magnitude and seeing
conditions. For the classification purpose we used MK standard
stars taken with the same configuration at CASLEO and the Digital
Spectra Classification of R.O. Gray ``Digital Spectra
Classification''
\footnote{http://nedwww.ipac.caltech.edu/level5/Gray} and ``MK
Standard Stars online''
spectra\footnote{http://stellar.phys.appstate.edu/Standards}. An
estimate by eye of the accuracy of our classification yields a
potential half a sub-type as maximum departure from the right
classification. Table 4 shows the MK spectral types obtained
together with photometric magnitudes and color indices for all
clusters but Ruprecht 60. Unfortunately, most of the
brightest stars in the field of this cluster are too faint ($V
\geq 13$ mag) to get an adequate compromise between the
information obtained and a reasonable time consumption.

\begin{center}
\begin{figure}[t!]
% \begin{flushleft}
% \fbox{}
\resizebox{13cm}{!}{\includegraphics{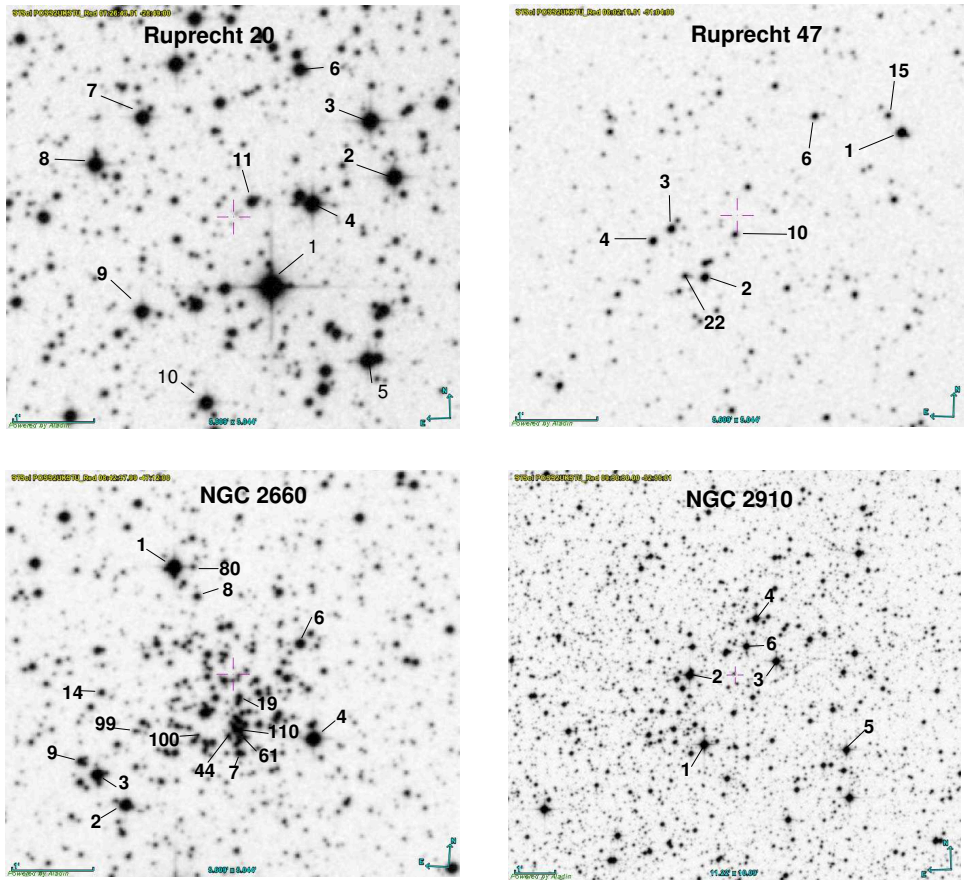}}
\caption{\small{Spectroscopic finding charts. Numbers are in the
notation used in this work (see Table 4). North and East as in
Fig. 1}} \label{cumulos_perfiles}
% \end{flushleft}
\end{figure}
\end{center}

\begin{center}
\begin{figure}[t!]
% \begin{flushleft}
% \fbox{}
\resizebox{13cm}{!}{\includegraphics{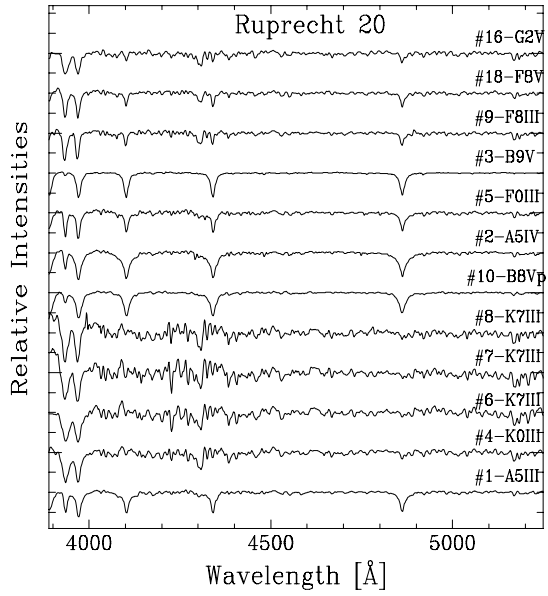}}
\caption{\small{Spectra of stars in Ruprecht 20. Numbers indicate
stars identified in Table 4.}} \label{cumulos_perfiles}
% \end{flushleft}
\end{figure}
\end{center}

\begin{center}
\begin{figure}[ht!]
% \begin{flushleft}
\resizebox{13cm}{!}{\includegraphics{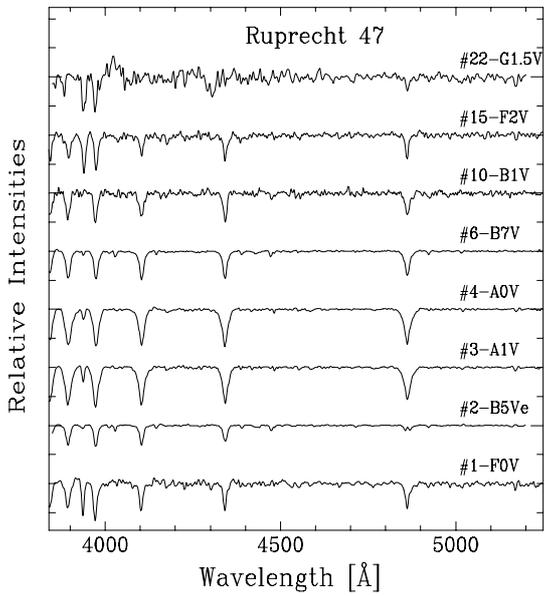}}
\caption{\small{Idem Fig. 3 for stars in Ruprecht 47}}
\label{cumulos_perfiles}
% \end{flushleft}
\end{figure}
\end{center}

\begin{center}
\begin{figure}[ht!]
% \begin{flushleft}
\resizebox{13cm}{!}{\includegraphics{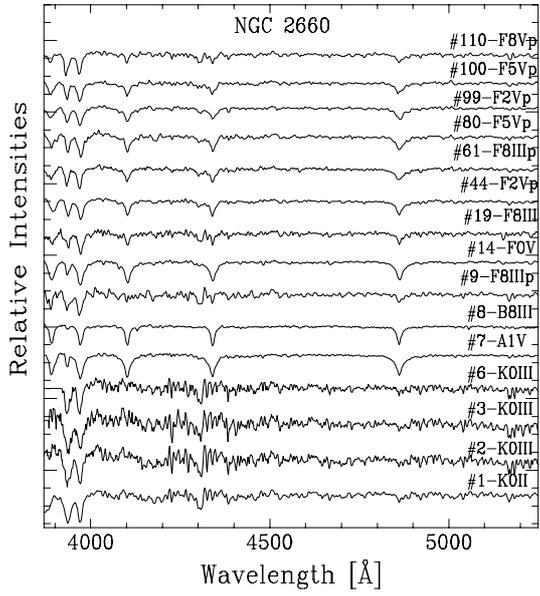}}\\
\caption{\small{Idem Fig. 3 for stars in NGC 2660 }}
\label{cumulos_perfiles}
% \end{flushleft}
\end{figure}
\end{center}

\begin{center}
\begin{figure}[ht!]
% \begin{flushleft}
% \fbox{}
\resizebox{13cm}{!}{\includegraphics{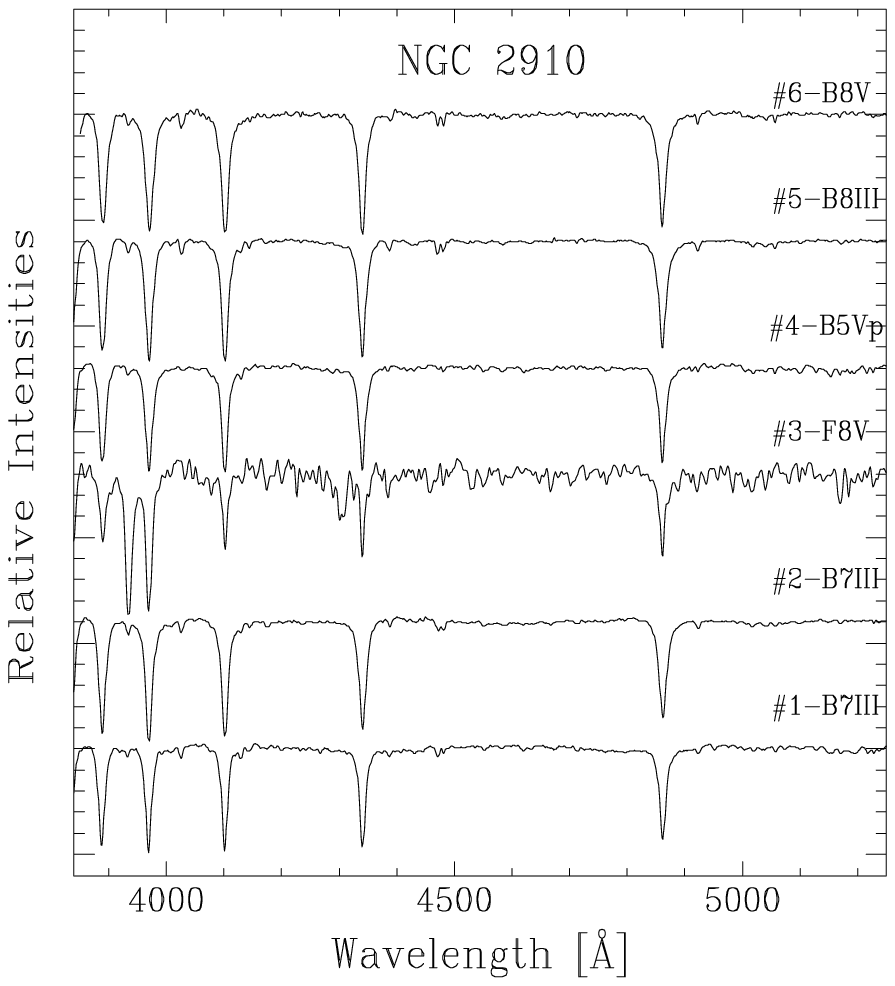}}
\caption{\small{Idem Fig. 3 for stars in NGC 2910}}
\label{cumulos_perfiles}
% \end{flushleft}
\end{figure}
\end{center}

Finally we want to mention that for a number of stars
discrepancies were found in the relative intensities of the K CaII
lines and the spectral types assigned according to the H line
intensities. This is, intensities of some Calcium lines belong to
spectral types that are not exactly congruent with spectral types
based in H lines. These peculiarities were found for stars in
Ruprecht 20 and a description of them are given at the bottom of
Table 4. Fig. 3 to 6 are the spectrograms of every stars we have
classified in our clusters. Numbers in the figures are in our own
notation indicated in Table 4.

\begin{table}
\begin{small}
\caption{MK spectral classification for stars in four of the five clusters.}
\begin{tabular}{l|l|c|lrrrrrr}
\hline
\hline
Cluster & Old ID & Our ID & MK ST & $V$ & $(B-V)$ & $(U-B)$ & $E_{B-V}$ & $M_{V}$ & $d_{\odot}$ [kpc]\\
\hline
\multirow{12}{*}{Ruprecht 20}
&VM72~ 1  & 1  & A5III &  8.79 & 0.28  &  0.12 & 0.13 &  0.70   & 0.34  \\
&VM72~ 2  & 4  & K0III & 11.60 & 1.03  &  0.78 & 0.03 &  0.70   & 1.45  \\
&VM72~ 3  & 6  & K7III & 11.64 & 1,57  &  1.76 & 0.04 & $-$0.30   & 2.28  \\
&VM72~ 4  & 7  & K7III & 11.67 & 1.58  &  1.73 & 0.05 & $-$0.30   & 2.05  \\
&VM72~ 5  & 8  & K7III & 11.91 & 1.31  &  1.16 & 0.00 & $-$0.30   & 2.77  \\
&VM72~ 6  & 10$^\ast$  & B8V    & 12.65 &  0.16 & 0.14 & 0.27 & $-$0.25   & 2.56  \\
&VM72~ 7  & 2  & A5IV  & 11,69 & 0.19  &  0.17 & 0.04 &  1.30   & 1.13  \\
&VM72~ 8  & 5  & F0III & 12.11 & 0.42  &  0.20 & 0.12 &  1.50   & 1.10  \\
&VM72~ 9  & 3  & B9V   & 12.04 & 0.04  & $-$0.07 & 0.11 &  0.20   & 1.99  \\
&VM72~ 10 & 9$^\ast$   & F8III & 12.39  & 0.56  &  0.10 & 0.00 &  1.30   & 1.60  \\
&VM72~ 11 & 18$^\ast$  & F8V   & 13.81  & 0.49  &  0.062 & 0.00 &  4.00   & 0.95  \\
&         & 16   &G2V  & 13.48 & 0.64  &  0.22 & 0.01 &  4.70   & 0.56  \\
\hline

\multirow{8}{*}{Ruprecht 47}
&VM72~ 1   &  4  &  A0V    & 12.01 & 0.22 &  0.10 &  0.24  &  0.65 & 1.33  \\
&VM72~ 2   &  3  &  A1V    & 11.92 & 0.26 &  0.13 &  0.27  &  1.00 & 1.04  \\
&VM72~ 3   &  2  &  B5Ve   & 11.52 & 0.17 & $-$0.46 &  0.34  & $-$1.20 & 2.16  \\
&          & 22  &  G1.5V  & 13.70 & 1.11 &  0.68 &  0.51  &  4.60 & 0.32  \\
&VM72~ 4   & 10  &  B1V    & 12.98 & 0.16 & $-$0.38 &  0.42  & $-$3.20 & 9.46  \\
&VM72~ 7   &  6  &  B7V    & 12.50 & 0.15 & $-$0.36 &  0.28  & $-$0.60 & 2.79  \\
&          & 15  &  F2V    & 13.41 & 0.50 &  0.08 &  0.15  &  3.60 & 0.74  \\
&VM72~  8  &  1  &  F0V    & 10.93 & 0.43 &  0.13 &  0.13  &  2.70 & 0.40  \\
\hline

\end{tabular}
\label{tab_spectra}
\newline

In Old ID column VM72 refers to Vogt \& Moffat (1972) numbering.\\
Ruprecht 20 $^\ast$: 10 KCaII(A1V), H(B8V) (broadened lines); 9
KCaII(F0), H(F8III); 18 KCaII(F8), H(F8), GB(F8), Mgab(F0)
\end{small}
\end{table}

\setcounter{table}{3}
\begin{table}
\begin{small}
\caption{(continued)}
\begin{tabular}{l|l|c|lrrrrrr}
\hline
\hline
Cluster & Old ID & Our ID & MK ST & $V$ & $(B-V)$ & $(U-B)$ & $E_{B-V}$ & $M_{V}$ & $d_{\odot}$ [kpc]\\
\hline
\multirow{6}{*}{NGC 2910}
&& 1 &  B7III   &  9.69 & 0.03 & $-$0.43 & 0.15 & $-$1.50 & 1.40  \\
&& 2 &  B7III   &  9.97 & 0.10 & $-$0.24 & 0.22 & $-$1.50 & 1.44  \\
&& 3 &  F8V     & 10.70 & 0.51 &  0.07 & 0.00 &  4.00 & 0.22  \\
&& 4 &  B5Vp    & 10.77 & 0.08 & $-$0.34 & 0.24 & $-$1.20 & 1.76  \\
&& 5 &  B8III   & 10.80 & 0.15 & $-$0.11 & 0.27 & $-$1.20 & 1.71  \\
&& 6 &  B8V     & 11.05 & 0.06 & $-$0.18 & 0.17 & $-$0.25 & 1.43  \\
\hline
\multirow{15}{*}{NGC 2660}
&& 1      &  K0II     & 10.75 & 1.26 &  0.98 & 0.18 & $-$2.30 & 3.15  \\
&& 2      &  K0III    & 12.17 & 1.38 &  1.33 & 0.38 &  0.70 & 1.14  \\
&& 3      &  K0III    & 12.18 & 1.72 &  1.56 & 0.72 &  0.70 & 0.71  \\
&& 6      &  K0III    & 13.17 & 1.27 &  0.95 & 0.27 &  0.70 & 2.12  \\
&& 7      &  A1V      & 13.91 & 0.47 &  0.39 & 0.49 &  0.65 & 2.23  \\
&& 8      &  B8III    & 13.91 & 0.32 & $-$0.16 & 0.43 & $-$1.20 & 5.69  \\
&& 9      &  F8III    & 13.97 & 1.05 &  0.57 & 0.51 &  1.20 & 1.73  \\
&& 14     &  F0V      & 14.36 & 0.53 &  0.32 & 0.21 &  2.70 & 1.59  \\
&& 19     &  F8V      & 14.40 & 1.26 &  0.93 & 0.33 &  4.00 & 0.75  \\
&& 44     &  F2Vp     & 14.85 & 0.68 &  0.28 & 0.33 &  3.00 & 1.46  \\
&& 61     &  F8IIIp   & 15.13 & 0.88 &  0.40 & 0.34 &  1.20 & 3.76  \\
&& 80     &  F6Vp     & 15.32 & 0.71 &  0.39 & 0.27 &  3.50 & 1.57  \\
&& 99     &  F2Vp     & 15.68 & 0.60 &  0.33 & 0.25 &  3.00 & 2.40  \\
&& 100    &  F6Vp     & 15.68 & 0.65 &  0.27 & 0.21 &  3.50 & 2.02  \\
&& 110    &  F8Vp     & 15.80 & 0.70 &  0.28 & 0.18 &  4.00 & 1.77  \\
\hline
\end{tabular}
\label{tab_spectra-cont}
\end{small}
\end{table}

\section{Getting the cluster fundamental parameters}

In the following subsections we describe our method to analyze
each cluster. Just to put the reader in the Galactic context we
show in Fig. 7 the location of each of the five clusters in this
study projected against a $\sim 40^\circ \times 40^\circ$ image of
the Vela Gum adapted from the SHASSA SM-1 survey. Table 5 is a
summary of fundamental parameters for all cluster as found in the
literature (last raw in each cluster is the result from this work
in advance).

\vspace{-3.372ex} %quita espacio vertical
\begin{table}[htb!]
% \scalebox{1}{
\caption{\textbf{Earlier and present study results} in the five clusters}
\begin{tabular}{c|c|ll|c|c|c}
\multicolumn{7}{c}{\begin{normalsize}Clusters parameters\end{normalsize}} \\
\hline \hline \multirow{2}{*}{Cluster} & $\alpha_{2000}$ &
\multicolumn{1}{c}{\multirow{2}{*}{Source}} & &
\multirow{2}{*}{$E_{(B-V)}$} &
\multirow{2}{*}{$d_{\odot}$(kpc)} & \multirow{2}{*}{Age (Gyr)} \\
& $\delta_{2000}$ & & & & & \\
\hline

\multirow{2}{*}{Ruprecht 20} &
\multirow{2}{*}{\begin{minipage}{0.6in}$7^{h} 26^{m} 43^{s}$\\
$-28^\circ 49^\prime$ \end{minipage}} & Vogt \& Moffat (1972)
& & 0.1 & 1.21 & - \\
& & This work & & - & - & - \\ \hline

\multirow{3}{*}{Ruprecht 47} &
\multirow{3}{*}{\begin{minipage}{0.6in}$8^{h} 02^{m} 19^{s}$\\
$-31^\circ 04^\prime$ \end{minipage}} &
Vogt \& Moffat (1972) & & 0.28 & 3.03 & - \\
& & Kharchenko et al. (2005) & & 0.25 & 3.01 & 0.11 \\
& & This work & & $0.32\pm0.02$ & $4.1\pm0.4$ & 0.06 - 0.08 \\
\hline

\multirow{2}{*}{Ruprecht 60} &
\multirow{2}{*}{\begin{minipage}{0.6in}$8^{h} 24^{m} 27^{s}$\\
$-47^\circ 13^\prime$ \end{minipage}} &
Bonatto \& Bica (2010) & & $0.64\pm0.1$ & $6.16\pm 0.88$ & $0.4\pm0.1$ \\
& & This work & & $0.37\pm0.05$ & $4.2\pm0.2$ & 0.8-1.0 \\ \hline %& $\sim[Fe/H]_{Solar}$

\multirow{8}{*}{NGC 2660} &
\multirow{8}{*}{\begin{minipage}{0.6in}$8^{h} 42^{m} 39^{s}$ \\
$-47^\circ 12^\prime$ \end{minipage}} &
Hartwick and Hesser (1973) & & $0.38\pm0.05$ & $2.88\pm0.01$ & 1.20 \\

&  & Lynga (1987) &  & 0.35 & 2.1 & 1.58 \\ %\cline{4-9} %& $[Fe/H]=0.06$
&  & Hesser \& Smith (1987) & & $0.35\pm0.03$ & 4.35 & - \\ %\small{Giants DDO photometry} %\cline{4-9} %& $[Fe/H]=-0.4$

&  & Frandsen et al. (1989) & & 0.36 & 2.9 & 1.2 \\ %\cline{4-9}

&  & Geisler et al. (1992) &   & $0.37\pm0.05$ & - & 1.7 \\ %& $[Fe/H]=-1.05\pm0.16$ %\cline{4-9}

&  & Friel (1995) & & - & 2.89 & 0.8 \\ %\cline{4-9}

&  & Sandrelli et al. (1999) & & 0.37-0.42 & 2.63-2.88 & $<1$ \\ %\cline{4-9} %& $\sim[Fe/H]_{Solar}$

&  & This work & & 0.33 & $2.7\pm0.2 $& $\sim1.58$ \\
\hline

\multirow{5}{*}{NGC 2910} &
\multirow{5}{*}{\begin{minipage}{0.6in}$9^{h} 30^{m} 22^{s}$\\
$-52^\circ 54^\prime$ \end{minipage}} & Becker (1960) &
& 0.09 & 1.25 & 0.04 \\
&  & Topaktas (1981) & & 0.05 & 1.32 & - \\
&  & Ramsay \& Pollaco (1992) & & $0.11\pm0.02$ & $1.45\pm0.01$ &
$<0.3$ \\
& & Kharchenko et al. (2005) &  & 0.34 & 2.61 & 0.03 \\
& & This work & & $0.22\pm0.05$ & $1.3\pm0.1$ & 0.06 \\
\hline
\end{tabular}
% } % cierra \scalebox
% \end{small}
\label{tab_param}
\end{table}

\begin{figure}[t!]
\centering
% \fbox{}
\resizebox{8cm}{!}{\includegraphics{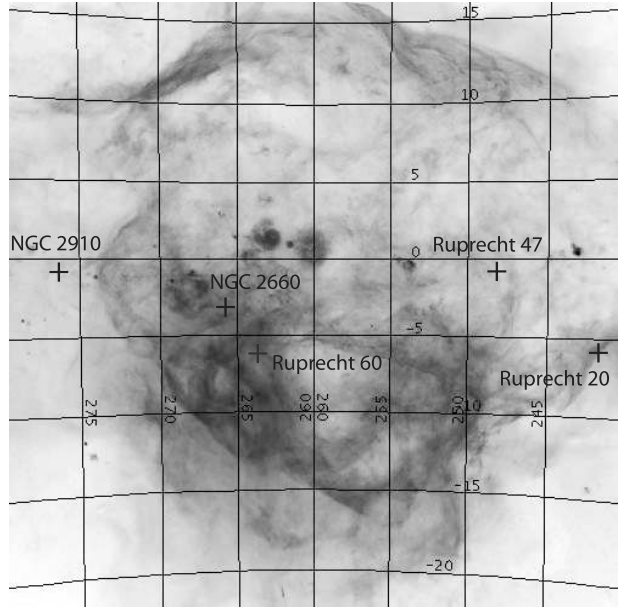}}
\caption{\small{The five clusters (plus symbols) in the present
sample projected against an H$\alpha$ full sky map image toward
the Vela Gum Nebula directions. Grid is in Galactic coordinates.}}
\label{}
\end{figure}

\subsection{Cluster sizes}

\begin{figure}[t!]
\centering
% \fbox{}
\resizebox{12cm}{!}{\includegraphics{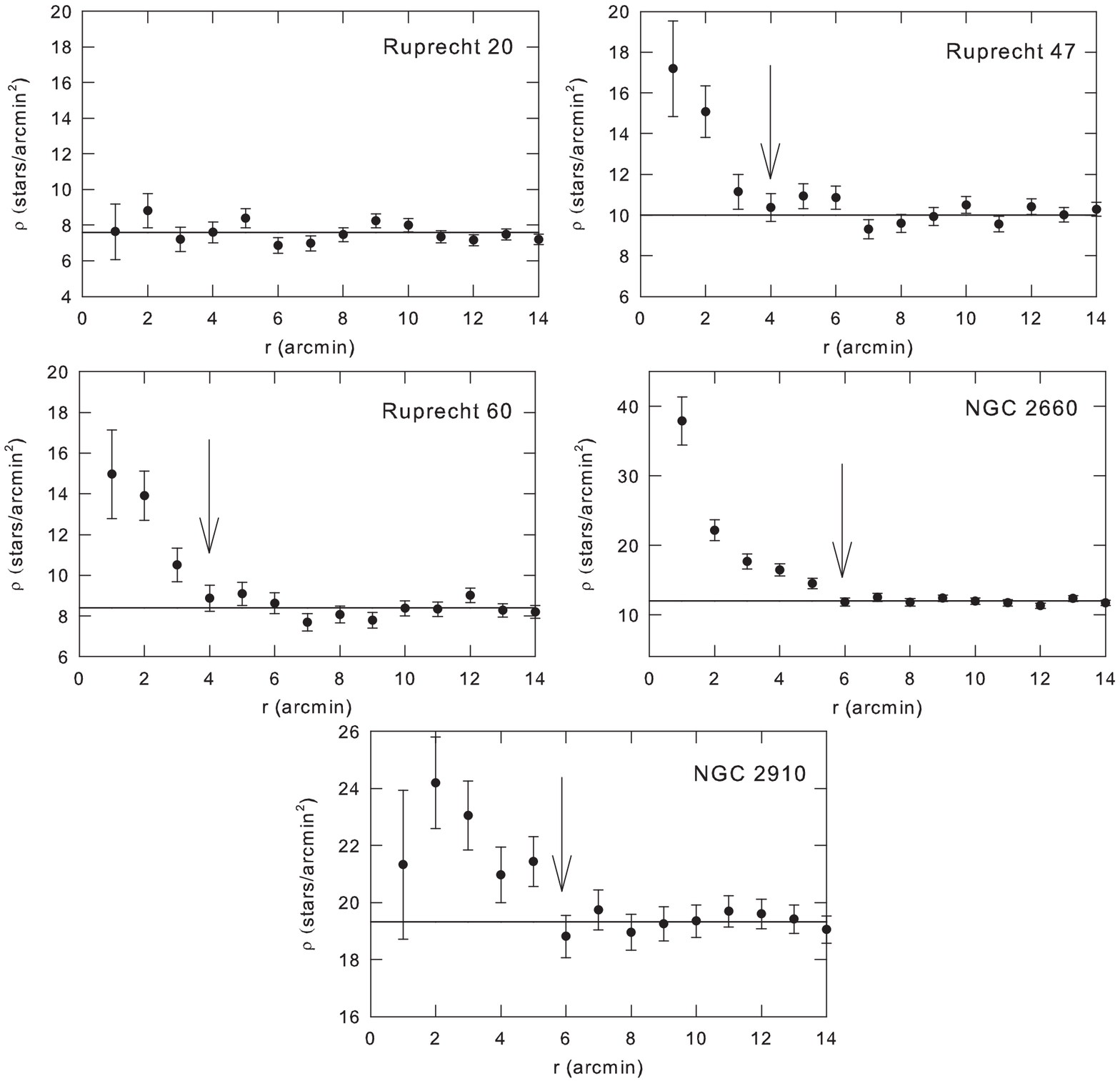}}
\caption{\small{Radial density profiles derived in the five
clusters with 2MASS data. The arrows indicate the point at which
the radial density merges with the background star density
\textbf{shown by solid lines}}} \label{}
\end{figure}

Accuracy in cluster parameters is achieved only when star
sequences sweeping a broad range of visual magnitudes are
identified and when contamination for field interlopers is
minimized. In this last aspect let us mention that cluster sizes
are routinely used as criteria for separating cluster dominated
from field-dominated regions in photometric diagrams (V\'azquez et
al. 2008). Although useful for recovering the precise locus
occupied by cluster members, it is worth recalling that the
cluster's linear size is a fundamental but overlooked quantity to
characterize these objects. In fact, precise sizes are necessary
for evaluating the cluster evolutionary status through analysis of
its density and other related parameters [see Aarseth (1996); de
La Fuente (1997); Kroupa et al. (2001)] such as the slope of the
mass spectrum. Usually, cluster sizes are computed by assuming a
centrally peaked spherical stellar distribution and then
determining the the distance at which its radial density profile
merges with the (flat) density of the stellar background. In
practice, this determination is normally done either by visually
setting this limit in a density radial plot or analytically by
fitting a King (1962) profile to an observed radial density
profile. In both approaches a spherically symmetric distribution
is assumed what may not be true since open clusters are, by
definition, irregular in shape. Moreover, in the case of the King
profiles, it is assumed that the system is dynamically relaxed
too. Since a good size estimation reduces quite considerably the
field star contamination, we restrict the analysis to stars within
the cluster limits.

Our starting point consists in adopting cluster center coordinates
given in available catalogs (e.g. WEBDA) and counting the number
of stars found in concentric rings $0.5^\prime$ in width
around them. Instead of using our own photometry we use
2MASS\footnote{http://www.ipac.caltech.edu/2mass/} data
(the three bands but particularly the deepest $K$ band)
that cover areas larger than ours. This way we ensure the
reliability of the cluster sizes and overcome the effects from
visual absorption. The star counts were divided by the area of
each ring following well established receipts (e.g. Carraro et al
2007). Fig. 8 shows density profiles for the 5 clusters in our
sample. Vertical error bars in the density profiles are computed
as the square root of the number of stars divided by the ring
surfaces. The radius of each cluster (see Table 1) is
shown by an arrows in the point where the cluster density merges
with the field stellar density level approximately (solid lines in
Fig. 8). In all the cases we estimated the field density level by
eye.

Finally, comparing Figs. 1 and 8 we want to emphasize that since
cluster radii are larger than the areas we have surveyed our
observations are entirely contained within the cluster boundaries
and therefore we do not expect high contamination by field
interlopers. Thus, we are confident that our observations are
useful to yield realistic cluster parameters.

\subsection{Memberships}

Open clusters are usually not far from the Galactic plane and
appear therefore projected against crowded stellar fields and
immersed in regions of high absorption. This makes the membership
assignment a difficult task, particularly amongst faint stars. All
across this article we address memberships applying the classical
method of checking simultaneously the consistency of the location
of each star in several photometric diagrams (two color diagrams,
hereinafter TCDs, and different color-magnitude diagrams, hereinafter
CMDs). This method
is an efficient tool when membership estimates rely on a careful
inspection of the TCD and consistent reddening solutions are
applied. Contamination for field interlopers is dramatic in
crowded fields and may lead, on a side, to bad membership
assignments and, on the other consequently, to wrong cluster
parameters. It happens the turn-off point position may be strongly
affected by interlopers even in the bright portion of the CMDs.
However, we use the $(U-B)$ index that is much more sensitive to
stellar temperature changes than the $(B-V)$ index (Meynet et al.
1993) so that the $V$ vs $(U-B)$ CMD is quite useful for
eliminating most of A-F-type field stars contaminating precisely
the vertical part of the cluster sequences, especially of young
clusters.

\subsection{Reddening, distances and ages}

The reddening value $E_{(B-V)}$ in each cluster has been computed
by superposition of the Schmidt-Kaler (1982) [SK82 hereinafter]
ZAMS in the TCD of the cluster. We shift the ZAMS along the
reddening line (given by the standard relation $E_{(U-B)} = 0.72
\times E_{(B-V)} + 0.05 \times E_{(B-V)}^2$) until the best fit to
the cluster sequence is achieved. On the other hand, the knowledge
of the ratio of total to selective absorption, $R \times
E_{(B-V)}=A_V$, must be well established to compute a confident
cluster distance. Though the analysis in Moitinho (2001) carried
out for several open clusters in the TGQ suggests that the
extinction law is normal, an additional check of the $R-$value was
performed in the $(B-V)\,vs\,(V-I)$ plane in Ruprecht 60, NGC 2660 and
NGC 2910. In this diagram, if the slope of the standard reddening
line $E_{(V-I)}/E_{(B-V)} = 1.244$ [Dean et al. (1978)] follows
the stellar distribution this fact means the extinction law is
$R=3.1$. Apparent cluster distance moduli were derived then by
superposing the SK82 ZAMS onto the observed cluster sequences and
the cluster distances from the Sun ($d_\odot$) are obtained after
removing absorption from apparent moduli. The fitting error in
each cluster was estimated by eye inspection.

As for cluster ages they were derived under the assumption
clusters are of solar metal content. We superposed isochrones
produced by Girardi et al. (2000) onto the cluster sequences (once
they have been well established) putting special emphasis in
fitting not only probable members in the upper sequence but also
covering the largest possible magnitude range. However, in four
out of the five clusters we were forced to use two ischrones to
establish an age range. Various factors such as remaining material
from the star formation processes (affecting young clusters),
undetected binary stars or unresolved double stars, altogether
scatter stars around the mean sequence causing difficulties to
achieve a good fit by a single isochrone. In older clusters these
difficulties are amplified because of the need to fit
simultaneously main sequence and red clump stars as we show in
subsections 4.3. and 4.4.

\subsection{Reddening and distances of stars with spectral types}

Spectral types are a powerful and independent tool to estimate
distances and color excesses of individual stars and become
indispensable to corroborate distances to open clusters. Combining
MK types and $UBV$ photometry we can easily get the distance of
individual stars via the well know spectroscopic parallax method.
In the cases of stars in Table 4 we proceed to assign intrinsic
$(B-V)_0$ and $(U-B)_0$ colors based in MK types following the
relations given in SK82. Intrinsic colors according to MK types
and observed colors allow to get the reddening $E_{(B-V)}$ and
derive the $A_V$ to produce free absorption magnitudes $V_0$ mag
using the expression given below. In a couple of late type stars
the reddening turned out to be negative. We think this is due to
the low reddening of these stars and uncertainties in the true
colors of evolved stars. In these cases we adopt a reddening value
$E_{(B-V)}=0.0$. Combining the MK types with absolute magnitudes
$M_V$ using SK82 relations we are able to get individual star
distances. The main sources of errors in distances come from
photometric uncertainties in color and magnitudes. This is, the
formula:

\begin{equation}
V-M_V = -5+5 \log{d}+A_V
\end{equation}

\noindent leads to an error in distance $\epsilon(d)$ of the form:

\begin{equation}
\epsilon(d)= \ln{10}\times d \times 0.2 [\sigma_V + 3.1
\sigma_{B-V}]
\end{equation}

\noindent where $\sigma_V$ and $\sigma_{B-V}$ are typical
photometric errors in our sample.

This procedure has been successfully applied in Perren et al.
(2012). In the current case, we assume no error in the extinction
law ($R=3.1$) and no error in the absolute magnitudes ($M_V$) of the
stars. The photometric errors of individual stars with spectral
types are all below 0.04 which means that errors in distances
derived with the spectroscopic parallax method are all below 10\%.
Table 4, columns 8, 9 and 10, contains the estimated $E_{(B-V)}$
excesses, adopted $M_V$ and distances for isolated stars with
spectral types.

\section{Cluster by cluster analysis}

\subsection{Ruprecht 20}

\begin{figure}[t!]
\centering
% \fbox{}
\resizebox{17cm}{!}{\includegraphics{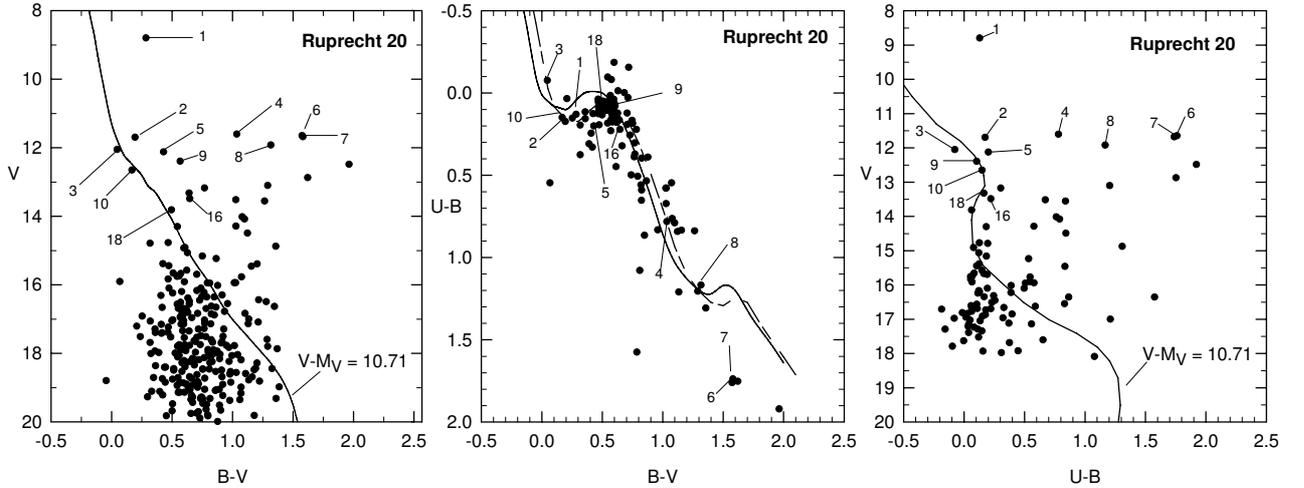}}
\caption{\small{From left to right: $V~vs~(B-V)$ CMD, $(U-B)~vs~(B-V)$
TCD and $V~vs~(U-B)$ CMD for Ruprecht 20. Stars with MK types are
pointed out. The SK82 ZAMS (dashed line) has been fitted to an
excess value $E_{(B-V)}= 0.1$ in the TCD. Solid lines in the CMDs
are the SK82 ZAMS fitted to an apparent distance modulus
$V-M_V=10.71$ as suggested by Vogt and Moffat (1972). See text for
details.}} \label{ru_20}
\end{figure}

Ruprecht 20 was studied by Vogt \& Moffat (1972) who obtained
$UBV$ photoelectric photometry for 11 stars. Although their work
calls into question the existence of a true cluster in this
region, they assume that stars number 10, 3 and 18 (our ID) [6, 9 and
11 in Vogt \& Moffat notation] are potential members of luminosity
class V (their assumed luminosity classes are confirmed by our
spectroscopy in Table 4 but we do not agree they are potential
members of Ruprecht 20). Vogt \& Moffat (1972) obtained the
following cluster parameters $E_{(B-V)}=0.1$, $V-M_{V}=11.04$ and
$d_\odot=1.21$ kpc in case it is a real entity. This cluster has
been subject also of a specific survey aimed at determining radial
velocity amongst red giants to secure memberships (Mermilliod et
al. 2008). These authors could find two red giant stars that they assumed likely cluster members.

According to our data the cluster radial density profile of
Ruprecht 20 shown in Fig. 8, upper left panel, does not suggest
the presence of an open cluster since no star grouping is evident.
Indeed, the density profile resembles the typical profile produced
by random stellar fluctuations.

Figure 9 includes from left to right the $V~vs~(B-V)$,
$(U-B)~vs~(B-V)$ and the $V~vs~(U-B)$ diagrams. In the TCD, middle
panel, we superpose the ZAMS shifted by $E_{(B-V)}= 0.1$ according
to Vogt \& Moffat (1972) and, simultaneously, we show the ZAMS
fitted to the apparent distance modulus $V-M_V= 10.73$
($d_{\odot}=1.21$ kpc) suggested by them too. It is simple to see
that no cluster sequence is accompanied by this ZAMS fitting or,
in other words, it is hard to perceive a cluster sequence in these
diagrams. Instead, we see a sparse handful of stars, some of them
affected by a similar amount of reddening, $E_{(B-V)}= 0.1$, but
no traces of any cluster. This is, exactly, what Fig. 8 indicates.

We find the respective distances for the 10 stars with MK spectral
types (see Table 4). According to their classifications, stars 10,
3 and 18 (our ID) are located at distances of approximately 2.55
kpc, 1.99 kpc and 0.95 kpc respectively, demonstrating beyond any
doubt they do not belong to any stellar aggregate. The distance of
stars 10 and 3 are of the order of the distance of star 6 ($\sim
2.30$ kpc). But this last is of late evolved type, a K7III, while
star 6 is a B8V and star 9 is a B9V.

In short, the CMDs and the TCD of Ruprecht 20 confirm that no
cluster exists in this location but we are dealing with a sparse
group of stars mostly of dwarf and giant-types. To mention is the
fact that above the second knee of the ZAMS there are some stars
that are hard to interpret. They are located along the path of the
reddening in the TCD and could be very reddened blue stars.
However, they are too faint. They could be sdB type stars but
since their average absolute magnitudes are $\sim +5$ and our
magnitude cut-off is $V = 17$ mag we do not expect more than a
tenths of sdB stars in our frames. Certainly, spectroscopy could
help to solve the nature of these objects.

After the above analysis we understand that evidences provided by
our photometry and spectral analysis are irrefutable and we
conclude that the open cluster Ruprecht 20 is not a physical
entity. In our opinion the connection of two red giant stars as
suggested by Mermilliod et al. (2008) with this object is
unrealistic.

\subsection{Ruprecht 47}

\begin{figure}[t!]
\centering
\resizebox{16cm}{!}{\includegraphics{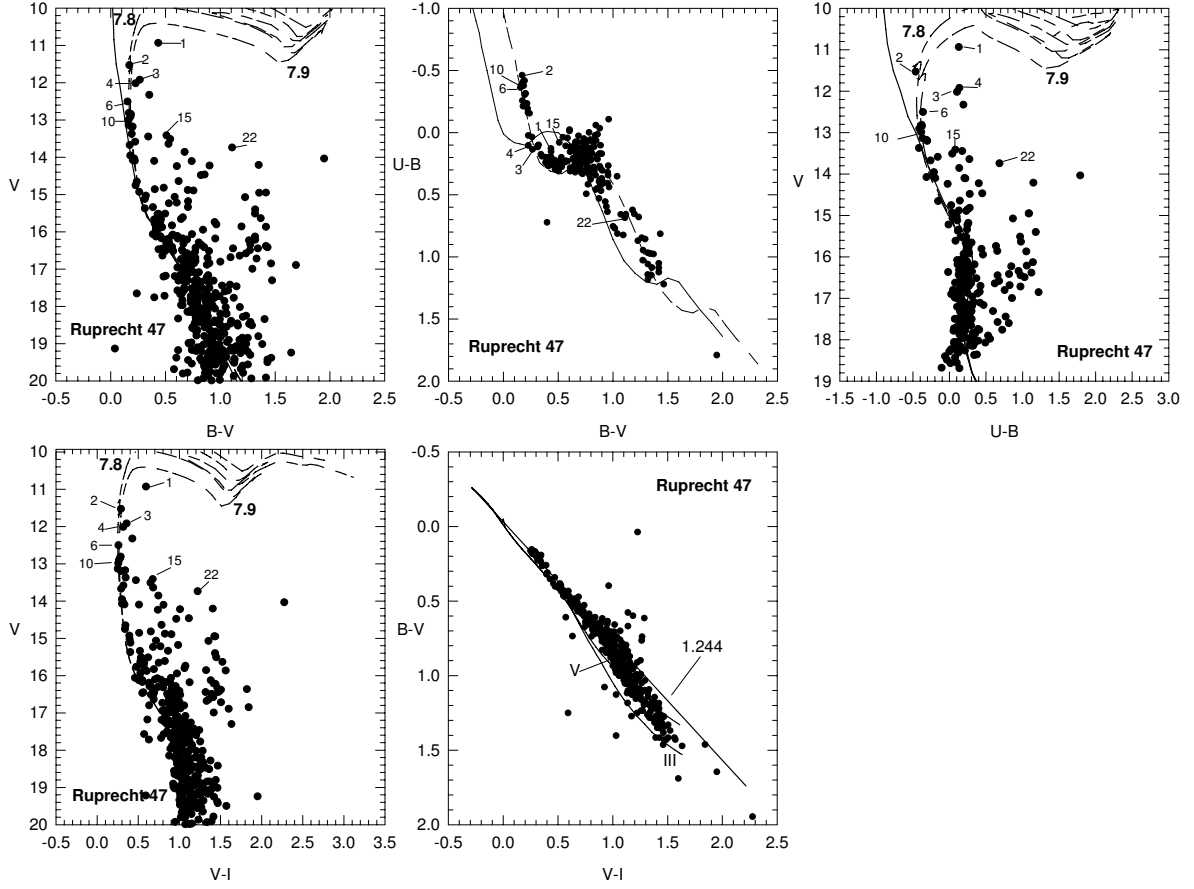}}
\caption{\small{Upper panels from left to right: $V~vs~(B-V)$ CMD,
$(U-B)~vs~(B-V)$ TCD and $V~vs~(U-B)$ CMD. Solid line in the TCD is the
SK82 ZAMS in its normal position while solid lines in the CMDs are
the SK82 ZAMS fitted to the cluster distance modulus. Dashed line
in the TCD is the ZAMS shifted by a reddening $E_{(B-V)}=0.32$.
The respective isochrone matchings in the CMDs are indicated by
dashed lines. Bold numbers by the isochrones are the $\log(t)$.
Left lower panel is the $V~vs~(V-I)$ CMD. Lines as in the upper
panels. Right lower panel is the color-color $(B-V)~vs~(V-I)$ diagram
including the intrinsic relations for stars of luminosity classes
V and III from Cousins (1978). The line with slope 1.244 is the
path of the reddening in this last plane and agrees with $R=3.1$.
Stars with MK spectral types are all pointed out, see text for
details.}} \label{r47-diagrams}
\end{figure}

Ruprecht 47 also was studied by Vogt \& Moffat (1972), who carried
out photoelectric photometry on 10 stars. Through these
observations they placed the cluster at a distance from the Sun of
3.03 kpc but, since their observations were limited to only a few
stars on the upper main sequence, the authors suggest the true
distance value can be quite different. Kharchenko et al. (2005)
also studied this cluster and found 7 probable members at a
distance of the order of 3 kpc with a reddening similar to the
obtained by Vogt \& Moffat (1972).

This case is quite different from the one of Ruprecht 20. In
principle, the radial density profile in Fig. 8 upper right panel
shows a clean star density profile extending up to 4$^\prime~$
where it merges with the background star density. This fact
suggests we are in the presence of a star overdensity.

Figure 10, the CMDs and TCD, show in turn a well defined star
sequence proper of a relatively young open cluster because of the
presence of several blue stars in the TCD (middle panel). Numbers
(our notation) in the diagrams point out stars with spectral
types. The superposition of the ZAMS in the TCD yields average
color excesses of $E_{(B-V)}=0.32\pm0.03$ and
$E_{(U-B)}=0.23\pm0.04$, a bit larger than earlier studies in
Table 5 but still, within 3$\sigma$, in agreement with Vogt \&
Moffat (1972) though quite far from the value found by Kharchenko
et al. (2005). In our interpretation, our reddening value is more
robust than previous ones since we can cover the cluster sequence
till the second knee in the TCD, the scatter of the data is low in
this diagram and the mean differences with Vogt \& Moffat (1972)
data are all quite acceptable.

The fitting of the SK82 ZAMS to the main sequence in the $V$ vs
$(B-V)$ and $V$ vs $(U-B)$ diagrams yielded an absorption free
distance modulus of $V_{0}-M_{V}=13.2\pm0.2$, corresponding to a
distance of $d_\odot =4.4\pm0.4$ kpc. This result is
quite different from earlier values determined by Vogt \& Moffat
(1972) and Kharchenko et al. (2005) for more that 1 kpc. So far as
we can interpret the data the explanation for such a difference
must be found in a simple fact already mentioned: our definition
of the cluster in the TCD is very good and, in addition, we can
track the cluster main sequence down to $V = 16$ mag in the CMDs.
This means a larger range of apparent magnitudes that allows to
achieve a better ZAMS superimposition and therefore more precision
in the cluster distance.

The cluster age was derived by matching the Girardi et al. (2000)
isochrones (dashed lines in the CMDs in Fig. 10). The best
possible matches were achieved for ages between 63 and 80 million
years ($7.8 \leq \log(t) \leq 7.9$). There is, as expected from
the difference in distance between this work and earlier ones, a
significant difference between our age value and the 117 million
years from Kharchenko et al. (2005). We think that the age
difference found is closely related with our ability to locate
precisely the cluster ``turn-off'' point at $V\sim 1.5$ mag. In
fact, that is why we differ from Kharchenko et al. (2005) who
derived the cluster parameters taking into account stars down to
$V\approx13$~mag. Certainly, since they had no opportunity to see
this point their distance and age were weakly determined.

Spectral types in the region of Ruprecht 47 include 8 bright stars
listed in Table 4 and pointed out in the corresponding panel of
Fig. 10. Three of them are of B-type, one, number 2 in our
notation, is an emission B5Ve star. Because of the small
$\beta$-value, Vogt \& Moffat (1972) suggested it is an emission
object that we confirm with our spectroscopy. Inspecting distances
shown in Table 4 we conclude that none of these 8 stars with
spectroscopic parallaxes have chances to become cluster members
since they are, on the average, twice closer to the Sun than the
cluster Ruprecht 47 situated at ($\approx $4 kpc). This
negative result is very interesting since without a spectroscopic
analysis most of the bright stars in this cluster would have been
assumed cluster members. The astrophysical impact of such an error
is clear; it produces an over-estimation of the number of massive
stars in the cluster head and severely modifies the cluster
initial mass function value (out of the scope of the our
investigation). That is why spectroscopy is so valuable. Now, it
is obvious the presence of other blue stars in the TCD of Ruprecht
47 and one may wonder how much field interlopers modify the
cluster parameter. We do not know the answer but we have been as
careful as possible in isolating the cluster region using star
counts. This said, we shall assume as likely cluster members the
other blue stars in Ruprecht 47 until the contrary is
demonstrated. Another interesting result from the of spectroscopic
parallaxes has to do with star 10 (in our notation). This is a
very early-type star, B1V, affected by an amount of reddening
comparable to the rest of stars. But quite surprisingly, this
object is a remote star located at almost 9.5 kpc from the Sun. If
our distance estimate is not wrong, this star becomes a serious
candidate to be member of the innermost part of the Outer Arm. We
discuss in section 5 the interpretation of this star in the
framework of the nearby Galaxy structure.

\subsection{Ruprecht 60}

Ruprecht 60 has been studied by Bonatto \& Bica (2010) through
2MASS photometry using a decontamination algorithm developed by
them. These authors stated the cluster is affected by a reddening
over $E_{(B-V)} = 0.6$ and placed it at a distance of more than 6
kpc. The age they found for Ruprecht 60 is $4\times 10^8$ yr.

This is the first time that $UBVI$ photometry is undertaken in
this overlooked object, probably, due to its weakness. In fact,
images from the \textit{Digital Sky Survey} (DSS) show an almost
irrelevant group of faint stars concentrated in a small portion of
the sky as seen in Fig. 1 (left middle panel). Notwithstanding,
our photometry that reaches stars as faint as $V\sim21$ mag is quite
concluding since the radial density profile shown in the middle
left panel in Fig. 8  confirms it is a true open cluster. On the
other hand, the extreme weakness of the brightest stars in this
object (they have magnitudes of $V\approx14.5$) precluded us from
taking spectra in a reasonable amount of observing time. So, our
analysis of this cluster rests upon photometric data alone.

\begin{figure}[t!]
\centering
\resizebox{17cm}{!}{\includegraphics{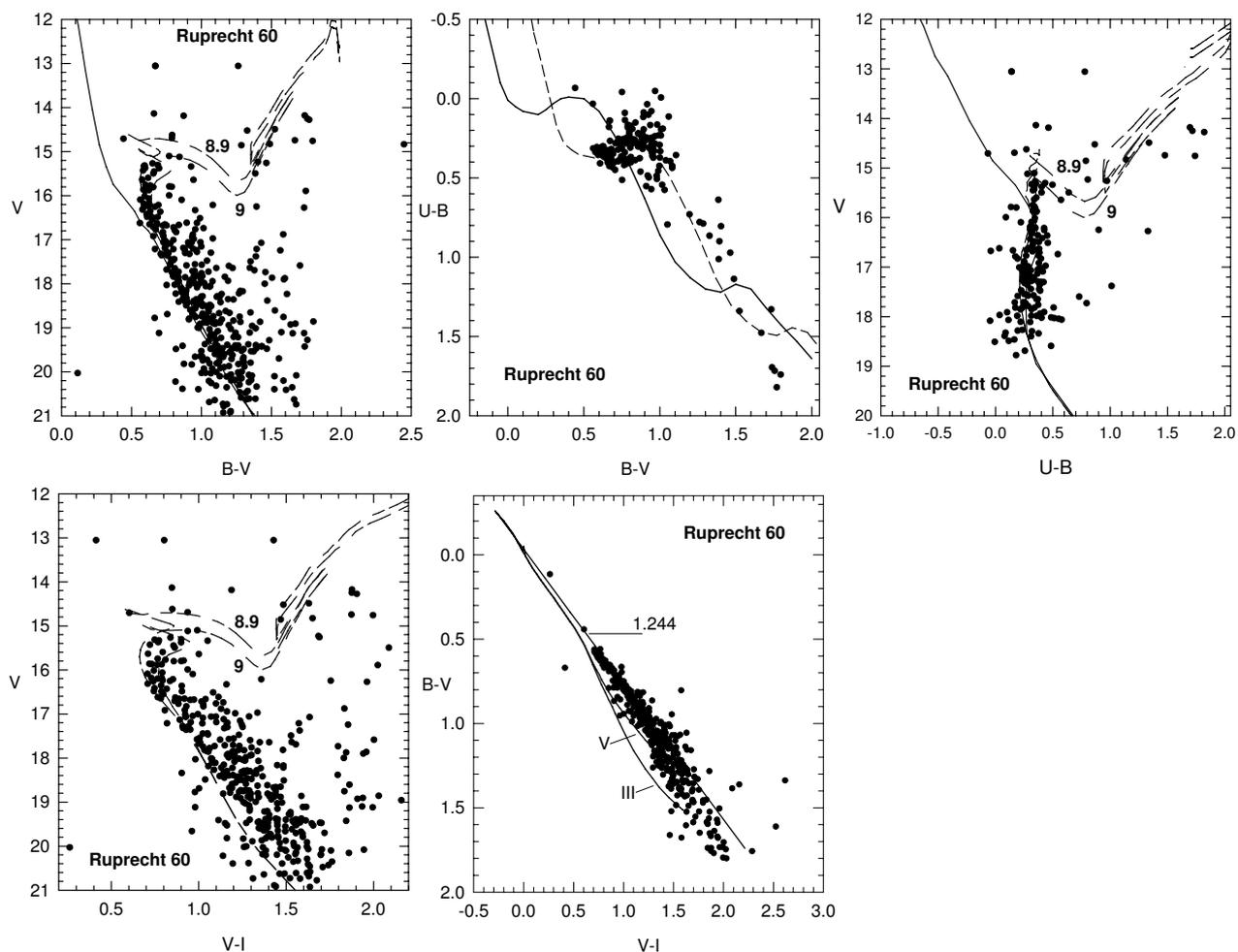}}
\caption{\small{Upper and lower panels as in Fig. 10. Idem for
symbols and lines. The ZAMS in the TCD is shifted by a reddening
$E_{(B-V)}=0.37$.}} \label{r60-diagrams}
\end{figure}

Figure 11 shows the CMDs $V~vs~(B-V)$, upper left panel, and the
$V~vs~U-B$, upper right panel, of Ruprecht 60. The middle upper
panel represents the corresponding TCD. In the lower left panel we
show the $V~vs~V-I$ CMD and in the lower right panel the
$(B-V)~vs~(V-I)$ color-color diagram. The $(U-B)~vs~(B-V)$ TCD
shows very few stars with $U-B$ inside the cluster area and
therefore with the highest chances to be physically related to the
cluster. We can fit these stars by shifting the intrinsic line by
$E_{(B-V)}=0.37\pm0.05$ and $E_{(U-B)}=0.27\pm0.05$. These values
are, in our interpretation of the TCD, the most probable color
excesses of Ruprecht 60. This is, we interpret there is no way the
color excess of this cluster is $E_{(B-V)}=0.64\pm0.1$ as claimed
by Bonatto \& Bica (2010). More interesting is the $V~vs~(B-V)$
CMD where a tight cluster main sequence dominates in the range 15
mag $<V<$ 19 mag. A worse isolation of this sequence is achieved
in the $V~vs~(V-I)$ CMD since the scatter of $(V-I)$ data is much
more pronounced. In all the CMDs, the $V~vs~(B-V)$ and
$V~vs~(V-I)$ particularly, the cluster sequence shows evidences of
evolution as stars leave the main sequence at $V\sim16.5$ mag.

Since the color excess of the cluster turned out to be
$E_{(B-V)}=0.37$ we compute the cluster distance by adjusting the
SK82 ZAMS in the CMDs $V~vs~(B-V)$ and $V~vs~(U-B)$ achieving the
best fit for an apparent distance modulus of
$V-M_{V}=14.25\pm0.10$. This corresponds to a true distance
modulus of $V_{0}-M_{V}=13.1$ that places Ruprecht 60 at a
distance of $d_{\odot}=4.2\pm0.2$ kpc. The absorption
law, that we use to get the absorption-free parameter above is
$R=3.1$ as evidenced by Fig. 11, lower right panel.

A precise estimation of the age of clusters with some degree of
evolution, as this is the case, is certainly complex in terms of
the morphology of the turn-off point. In fact, as summarized in
Friel (1995), the number of binary stars and blue stragglers in
the field of old open clusters has important implications in
interpreting the color-magnitude diagram and their comparison with
theoretical models, since the morphology of the turn-off region is
strongly affected by the presence of composite systems. When
binary systems evolve through the turn-off region of single/simple
stars, a color enlargement of the CMD diagram occurs near this
point and the systems move toward larger luminosity than that of
the true turn-off, before evolving into the region of red giants.
Apart from this, many old open clusters show stars in the
blue straggler region of the color-magnitude diagram, above and to
the blue of the main sequence turnoff. Radial velocity studies
indicate that a significant number of stars found in this region
are members of clusters and, among them, the spectroscopic binary
frequency exceeds 40\% [Milone (1992), Friel (1995)]. The
intersection of the evolutionary paths of single stars and binary
systems creates thus a more complex color and luminosity
distribution precisely in the region near the turn-off, where
clarity is necessary in order to adjust the set of isochrones and
determine the age based on the position of this point. In Ruprecht
60 CMDs we do not see a potentially large number of candidates to
blue straggler if any. As for the presence of binary systems there
is nothing we can say since our observations were not designed for
detecting this type of objects. Therefore, and under the
hypothesis that this cluster is solar metal content, the best fit
for the set of isochrones by Girardi et al. (2000) yielded, after
several tries, an age range between 800 and 1000 million years
($8.9 \leq \log(t) \leq 9$) as shown in Fig. 11.

Our data analysis yielded results quite different from Bonatto \&
Bica (2010). In fact, they have found, as seen in Table 5, a
larger reddening and a larger distance (almost 50\%) than ours.
And, in terms of the cluster age, ours is larger by a factor of
more than 2. Even though these authors applied a less personal
method (more sophisticated too) to estimate cluster parameters, we
interpret that these differences (cluster size aside) come from
the use of $JHK$  2MASS photometry. That is, getting the  cluster
reddening using $IR$ photometry may lead to wrong results if one
only rest on this photometry alone. As for the cluster age, the
isochrone matching shown in their Fig. 5 is somewhat weak mainly
because the cluster main sequence displays only 2 mag in $J$. On
the contrary, our optical CMDs show a more robust fitting not only
because of the larger magnitude range but also because our
isochrone matching follows pretty well the bend of the cluster
main sequence.

\subsection{NGC 2660}

\begin{center}
\begin{figure}[t!]
\centering
\resizebox{17cm}{!}{\includegraphics{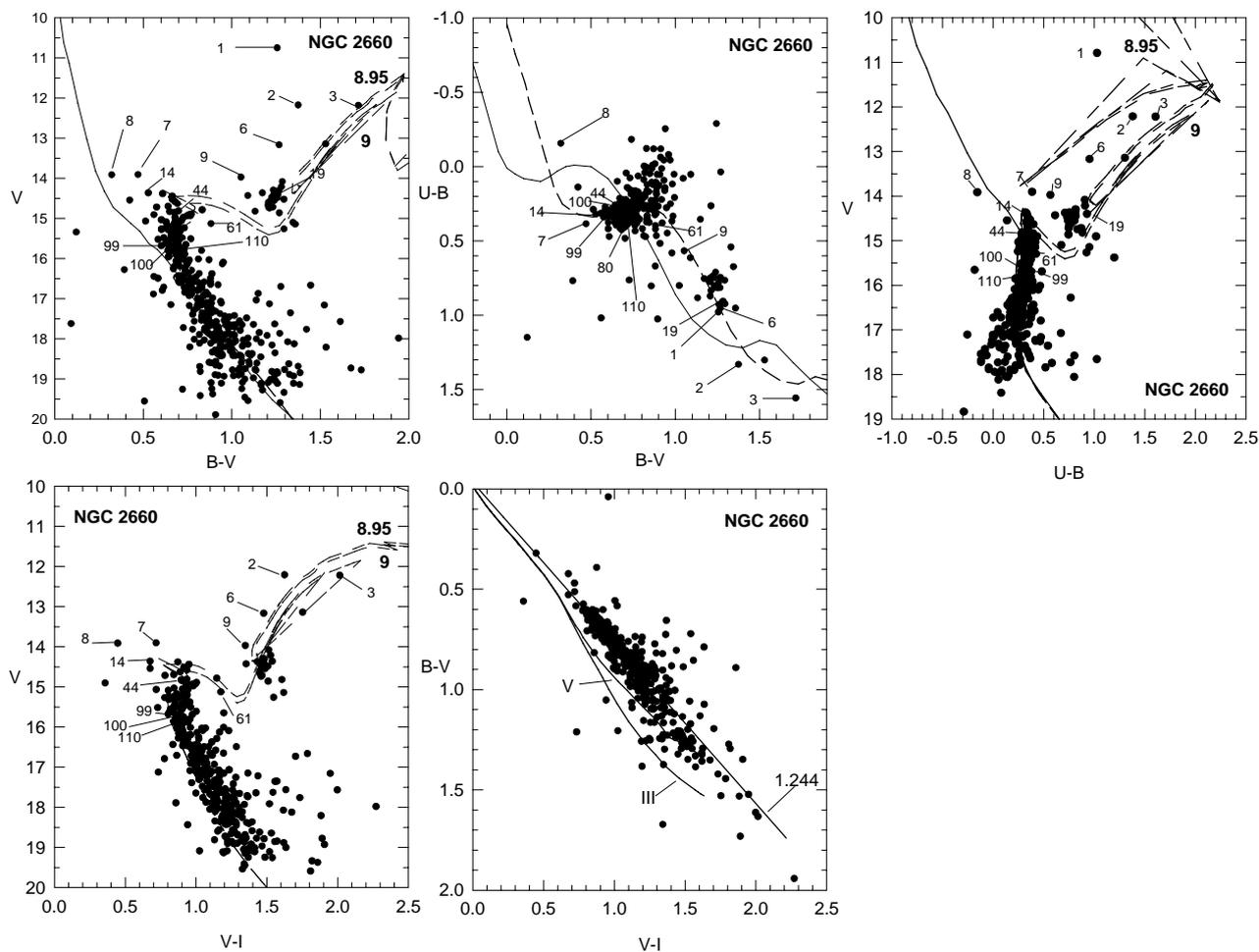}}
\caption{\small{Upper and lower panels as in Fig. 10. Idem for
symbols and lines. The ZAMS in the TCD is shifted by a reddening
$E_{(B-V)}=0.33$.}} \label{2660-hr}
\end{figure}
\end{center}

The first detailed study of NGC 2660 was carried out by Hartwick
\& Hesser (1973) by means of $UBV$, $uvby$ and H$\beta$
photoelectric photometry and $BV$ photographic photometry. Since
then several authors have studied this cluster as reported in
Table 5. It is noticeable that through the years different authors
with different techniques converge to a same $E_{(B-V)}$ color
excess and the distance within the usual errors. The exception is
the very discrepant value of $d_\odot =4.34$ kpc found by Hesser
\& Smith (1987) well above the mean of $\sim$ 2.8 kpc. The
situation is quite different in relation to the age of NGC 2660
that spans a range going from 0.7 to 1.7 Gyr from one author to
another. Certainly, the critical point resides in the metal
content of the cluster, a question that still remains open. Since
the purpose of our investigation excludes investigation of the
metal content in NGC 2660 we rest onto the recent papers of
Sestito et al. (2006) and Bragaglia et al. (2008) that concluded
this cluster is solar metal content.

Just for the sake of completeness we want to mention the most
recent paper of Sandrelli et al. (1999) based on $UBVI$ photometry
who found a probable 30\% population of binary stars in NGC 2660.
M{\o}lholt et al. (2006) detected a number of $\delta$ Scuti
variable stars in this object. Mermilliod et al. (2008), in turn,
included this cluster in their large survey for red giants in open
clusters identifying six members.

The radial density profile of NGC 2660, right middle panel in Fig.
8, shows a strong star density covering a radius of 6 arcmin
(2MASS data) that emerges above a quite flat star field. Our
survey is therefore covering the very nuclear cluster zone. In
Fig. 12 we show the CMDs $V~vs~(B-V)$, upper left panel, and the
$V~vs~(U-B)$, upper right panel. The middle upper panel shows the
$(U-B)~vs~(B-V)$ TCD. The lower left panel is the $V~vs~(V-I)$ CMD
and the lower right panel is the $(B-V)~vs~(V-I)$ color-color
diagram.

The TCD, like in Ruprecht 60, is very useful to give a reasonable
approximation to the true color excess affecting NGC 2660, even if
there are no blue-type stars. The best ZAMS shifting in the TCD is
for $E_{(B-V)}=0.33\pm0.03$, a value slightly lesser than previous
estimates reported in Table 5 but still in good agreement with the
historical mean for NGC 2660. The CMDs show a very well populated
cluster main sequence with a well defined turn-off. The data
scatter along the main sequence seen in the $V~vs~B-V$ and $V~vs~(V-I)$
is relatively little what suggests, in our opinion, that
contamination for field interlopers is low.

In the $V~vs~(B-V)$ and $V~vs~(V-I)$ CMDs of Fig. 8 we show the
ZAMS of SK82 fitted with an apparent distance modulus $V-M_V =
13.25\pm0.15$ and the above mentioned $E_{(B-V)}$. Since the
$(B-V)~vs~(V-I)$ color-color diagram (lower right panel in Fig. 8)
confirms that the absorption law $R=3.1$ is completely valid in
this region the absorption free distance modulus is
$V_0-M_V=12.2\pm0.15$ that places NGC 2660 at a distance
$d_\odot= 2.7 \pm0.25$ kpc. This value agrees with those found in
the literature (see Table 5).

A dominant feature in all the CMDs of NGC 2660 is the compact red
clump at $V\sim 14.5$, also noticeable in the TCD at $(B-V)
\sim1.25$ and $(U-B)\sim 0.8$. Adopting the same temperament as in
the case of the turn-off point of Ruprecht 60 we proceed to
estimate the age of the cluster looking for the best set of
isochrones embracing not only the turn-off point stars but also
the evident red clump. Again, assuming the cluster's metallicity
is solar, and by making use of Girardi et al. (2000) set of
isochrones, we find that the best fit corresponds to the age
interval $8.95 \leq \log(t) \leq 9$. Particularly good is the
isochrones fitting in the $V~vs~(U-B)$ CMD. Notwithstanding, the red
clump of this cluster raises some controversy since we could only
fit simultaneously the cluster turn-off and the red clump stars in
the $V~vs~(V-I)$ CMD. In the other two CMDs we note the clump looks
blue-shifted and becomes impossible to fit for an isochrone. This
effect has been noticed too in earlier works and that raised the
controversy about the metal content of stars in this cluster.
Photometry is unable to resolve the controversy. Just for the sake
of curiosity we have made several attempts using different metal
content isochrones without achieving a satisfactory result.
Clearly a profound spectroscopic analysis is needed in NGC 2660.

MK types for stars in this cluster reported in Table 4 are
illustrative in terms of comparing reddening and distance derived
from photometry with the same data from spectroscopy alone. Data
in Table 4 shows $E_{(B-V)}$ values ranging from 0.18 to 0.72
(only one star with this extreme value) while most stars
have reddening values near the mean of 0.33. Looking at individual
star distances with MK types in NGC 2660, we conclude that most of
them are in front of the cluster at more than 0.5 kpc from it.
These stars are 2, 3, 6, 9, 14, 19, 44, 80, 100 and 110.
Photometric arguments (the location of these stars in the
photometric diagrams in Fig. 12) favor that a number of them are
not members directly or have low chances to be related to the
cluster. Two stars, 8 and 61 are farther than the cluster
distance; the star 61 is otherwise bad located in the photometric
diagrams. We draw the attention to star 8, a B8III placed at a
distance $d_\odot \sim 5.7$ kpc. We will return to this star in
the next section. Star 1 has a chance, from a spectroscopic point
of view, to be a likely member; however, its position in
the photometric diagrams excludes it. Stars 99 and 7 are
relatively close to the cluster distance. In the case of star 7,
and given its location near the turn-off, it becomes a potential
blue straggler star. Like in Ruprecht 47 a large number of
stars with MK types turned out to be non members. This fact
permits us to throw away field interlopers with confidence and get
clean diagrams of the cluster. There is no influence of
interlopers on the cluster parameters since these latter depend on
the turn-off location and the red clump. There is no influence of
interlopers on the cluster parameters since these latter rely on
the location of the turn-off and the red clump.

\subsection{NGC 2910}

\begin{figure}[h!]
\centering
\resizebox{17cm}{!}{\includegraphics{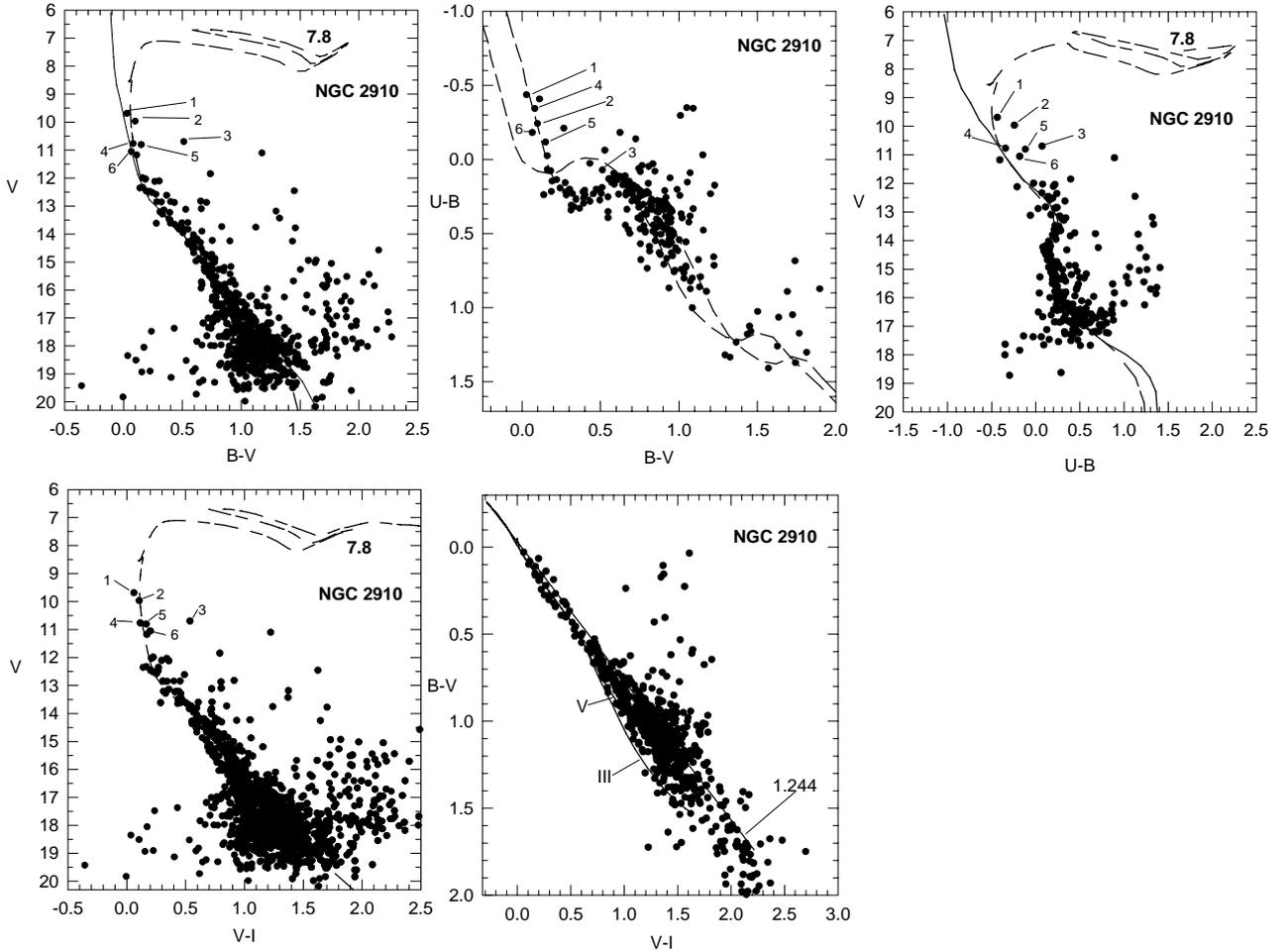}}
\caption{\small {Upper and lower panels as in Fig. 10. Idem for
symbols and lines. The ZAMS in the TCD is shifted by a reddening
$E_{(B-V)}=0.22$.}} \label{ngc2910-3-diagrams}
\end{figure}

Available information in the literature suggests NGC 2910 is a
young open cluster. It has been previously studied by Becker
(1960), Topaktas (1981), Ramsay \& Pollaco (1992) and Kharchenko
et al. (2005). Becker and Topaktas made use of the same data set
(the Becker (1960) one) founding a large number of members amongst
the bright stars in the zone. However, the deeper CCD photometric
survey carried out by Ramsay \& Pollaco (1992, 1994) could only
find 12 members. Kharchenko et al. (2005) revisited this cluster
finding only 9 probable members. Evidently there are difficulties
in understanding the upper main sequence of NGC 2910 that could be
easily solved by undertaking a vigorous spectroscopic program.
Although we have few stars with spectroscopy in this cluster we
hope we can improve the knowledge of the region as we show below.

Despite our photometric survey includes a large portion of the
cluster surface it is still insufficient to cover the entire
cluster given its radial extension (see Fig. 4 lower panel).
Photometric diagrams in Fig. 13 show an extended cluster main
sequence down to $V\sim 17$ mag with very low scatter around the
mean location. Contamination by field interlopers is not dramatic
and the cluster appears clean against the star background. As for
the cluster main sequence it appears affected by a uniform
reddening value that we estimate $E_{(B-V)}=0.22\pm 0.01$ and show
in Fig. 13 middle upper panel. This value disagrees with previous
reports found in the literature and indicated in Table 5. Our
reddening is more than twice the found for some authors but
Kharchenko et al. (2005) who found $E_{(B-V)}=0.34$ quite
unrealistic from the TCD in the middle upper panel in Fig. 13.

We applied the same procedures used in the other clusters in this
sample to derive the corresponding distance. The superposition of
the SK82 ZAMS in the $V~vs~(B-V)$ and $V~vs~(U-B)$ CMDs has been done
for a distance modulus $V-M_V= 11.22\pm0.15$. After correcting for
visual absorption, taking into account that the extinction law is
the standard one (see Fig. 13, lower right panel), the distance of
NGC 2910 from the Sun turns out to be $1.3\pm0.1$ kpc a coincident
value with earlier findings. Again we have a large discrepancy
with Kharchenko et al. (2005) who found 2.6 kpc.

As for the age of the cluster we find a single isochrone able to
fit the entire main sequence. This is the $\log(t)=7.8$ for an age
of 63 Myr. That is, from our data analysis, the cluster is a little
bit older than earlier estimates.

MK types for NGC 2910 in Table 4 leave only three stars with
chances to be cluster members; they are stars 1 and 2 of evolved
B-Types and star 6, another B-type star. The other two stars, 4
and 5 in our notation, are not members of the cluster. They are
too of B-types, one of late B-type and the other is a peculiar
B-type star. The remaining star with MK-types in NGC 2910 is a
background F-type star.

\subsection{Reddening path}

\begin{figure}[h!]
\centering
\resizebox{12cm}{!}{\includegraphics{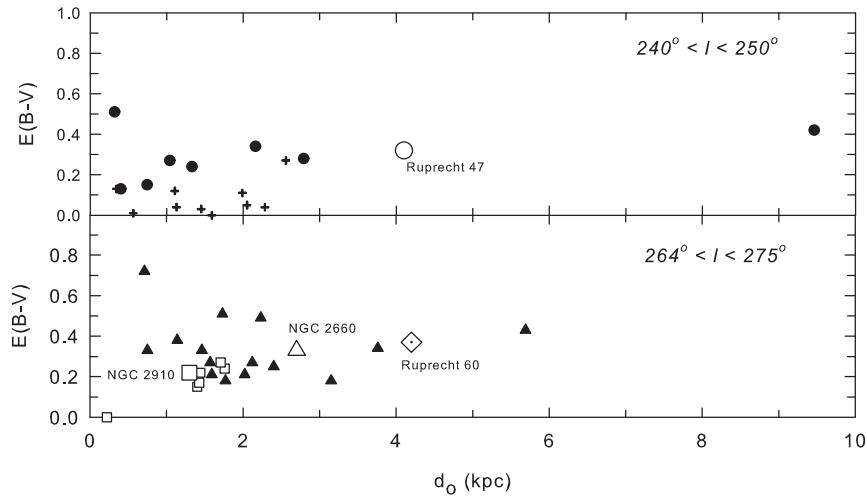}}
\caption{\small{The path of $E_{(B-V)}$ against distance from the
Sun. The upper panel includes stars in Ruprecht 20, crosses, stars
in Ruprecht 47 are indicated by small dots, large open circle
gives the position of the cluster Ruprecht 47. Lower panel
includes NGC 2910, big open square; small open squares indicate
stars in the area of NGC 2910. Triangles are for stars in NGC 2660
while the big open triangle is the cluster NGC 2660. Open diamond
gives the location of Ruprecht 60.}} \label{}
\end{figure}

We finally show in Fig. 14 the path of $E_{(B-V)}$ against the
distance for all the stars and clusters in the present sample. The
upper panel includes stars with MK types in the field of clusters
Ruprecht 20 and in Ruprecht 47. The position of Ruprecht 47 is
indicated. In the lower panel we do the same for Ruprecht 60 NGC
2660 and NGC 2910. Following well known relations that state that
$E_{(b-y)} = 0.7 E_{(B-V)}$ we compare our reddening estimations
with the findings of Kaltcheva \& Hilditch (2000). For both
regions under analysis, $240^\circ \leq l \leq 250^\circ$ and
$264^\circ \leq l \leq 275^\circ$ the reddening values are
entirely in agreement. The presence of some patches of dust may be
responsible for the couple of nearby stars in Fig. 14, in both
zones, showing high reddening values. Ruprecht 20 and Ruprecht 47
stars show different reddening behavior probably due to the fact
that Ruprecht 47 is located along the Galactic plane (steadily
increasing) while Ruprecht 20 is at more than 5 degrees below the
plane (quite flat).

\section{The Galaxy structure toward the Vela Gum}

One of the clusters in our sample is Ruprecht 20 that we
have already shown it is not a physical entity. This object is
projected against the north of the Vela Gum in a region also known
as Puppis association. The other cluster in our survey also in
Puppis is Ruprecht 47 (in the limit with Pyxis constellation).
Puppis is an intricate zone with a stellar structure hard to
elucidate. Humphrey (1978) found various components of what is
designated as Puppis OB1 association at 2.5 kpc from the Sun,
while Havlen (1976) had, a little before, reported a second, more
distant association, Puppis OB2 at 4.3 kpc. However, the
existence of these two separate associations have been questioned
by Kaltcheva \& Hilditch (2000), who could not find evidences of
them. Actually, these authors suggested the presence in
the same Galaxy direction of two other star groups (at 1 kpc the
first, and 3.2 kpc the second), but significantly lower in
latitude, at $b\sim-4.0^\circ$. These two young star groups would
surround, in projection, the star cluster NGC 2439,
located at 3.5-4.5 kpc from the Sun (see Fig. 6 in Kaltcheva \&
Hilditch (2000)). Ruprecht 20 is situated a little below
this last location and we found two stars, numbers 10 and 3 in our
notation, of B8V and B9V spectral types in the field of this
object at distances of $\sim$ 2.5 and $\sim$ 2.0 kpc
respectively. Within distance errors of the order of
15-20\% as indicated in subsection 3.3 these stars are
placed in front of the stellar aggregate 2 proposed by Kaltcheva
\& Hilditch (2000) making this stellar group a long structure
extending from 2 to 4 kpc from the Sun.

Ruprecht 47 is placed well along the Galactic plane at a distance
$d_\odot = 4.4$ kpc. It becomes therefore clearly coincident with
Puppis OB2 at 4.3 kpc according to Havlen (1976) assertion
and the genetic connection with Puppis OB2 is supported by its age
(less than 100 Myr). It is to mention that not far from the
position of Ruprecht 47 there exists the open cluster Haffner 20
($246.97^\circ$,~$-0.93^\circ$) with similar age (Carraro et al. 2015)
and at a distance of 5.5 kpc in the conjunction with the Perseus Arm. It
seems, therefore, that Puppis OB2 may extend farther, even much
more than previously indicated in the literature. Table 4 informs
as well the existence of two B-type stars in the field of Ruprecht
47, one B1V and the other B5Ve at distances of 2.0 and 2.2 kpc
respectively. If we follow the Humphreys (1978) discussion these
two stars belong to Puppis OB1. Connecting our findings with those
of Kaltcheva \& Hilditch (2000) it seems that early type stars in
Puppis constellation, including Puppis OB1 and OB2 and regions 1
and 2 from Kaltcheva \& Hilditch compose a sort of thickening
(maybe a bridge) of the Galactic disk of about 300 pc at
2-3 kpc from the Sun extending outward up to 6 kpc.

In the background of Ruprecht 47 there is another star, number 10
in our notation and classified B1V, placed at the huge distance of
$d_\odot = 9.5$ kpc. This is a surprising finding since this
distance implies we have detected a young stellar component of the
innermost side of the Outer Arm. This star is very distant no
matter the data set adopted. Indeed if we compute its
spectroscopic parallax with data from Vogt \& Moffat (1972) the
distance from the Sun would be of the order of 10 kpc.

NGC 2660 and NGC 2910 are two clusters placed in the Vel OB1
association. According to Humphreys (1978) the nucleus of the
association is placed at 1.8 kpc from the Sun. The cluster NGC
2660 is not subject of a discussion of this type in terms of its
belonging to Vel OB1 since the age, around 1 Gyr, excludes it.
However, there is another star, 8 in our notation, classified
B8III and placed at $\sim 5.7$ kpc from the Sun. We
discuss below the meaning of this star. As for NGC 2910, its
distance of 1.3 kpc makes it a sure member of the Vel OB1
association and therefore a member of the Local Arm. In the
cluster field there are also two stars, 4 and 5, B5Vp and B7III,
at $\sim 1.7$ kpc from the Sun respectively, that are likely
members of the Vela OB1 association.

\begin{figure}[h!]
\centering
\resizebox{9cm}{!}{\includegraphics{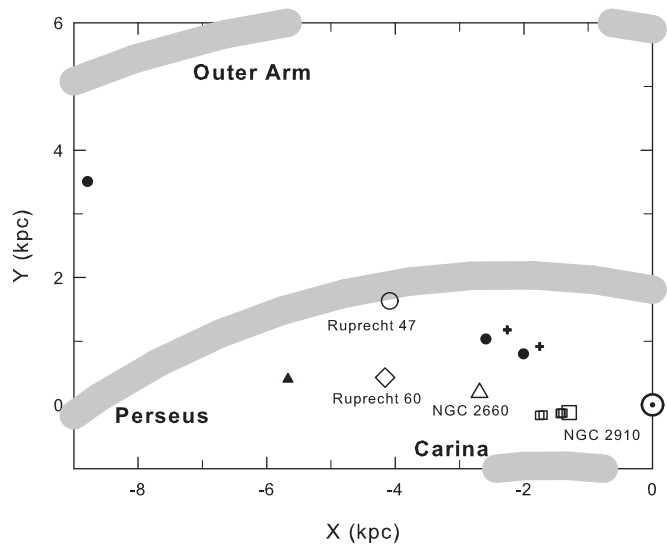}}
\caption{\small{X-Y projection onto the TGQ of blue stars and
clusters in the present study. Symbols as in Fig. 14. The Sun is
at 0,0 (circle with a dot inside). The trace of the Perseus,
Carina and Outer Arms (Vall\'ee (2005) are shown. The scale
indicates the Galactic longitudes.}} \label{}
\end{figure}

Figure 15 is the X-Y projection of the TGQ showing the
position of clusters and young stars analyzed in the present
study. The picture in Fig. 15 is complementary of similar ones we
have recently shown in Carraro et al. (2015) and enlarges the
available information by including more clusters and stars in
Puppis and Vela associations. Our Fig. 15 and Figs. 1 and 12 in
Carraro et al. (2015) show the presence of a large number of blue
stars and several young clusters composing a sort of stellar
bridge between the Perseus Arm and the Sun position covering from
$\sim 220^\circ$ to $\sim 275^\circ$ in Galactic longitude. That
is, the young stellar population we have found with spectroscopy
forms part of the Local Arm in the TGQ, strictly speaking in Vela
and Puppis associations. NGC 2910 is a young object member of the
Local Arm and close to the border of the third and fourth Galactic
quadrants. Probably this cluster is one of the farthest member of
the Local Arm in the direction to Vela. The cluster Ruprecht 47 is
also young and is placed close to the formal Galactic plane. So,
we understand that this object is a likely member of the Perseus
Arm (or a very distant member of Puppis OB2 in the Local Arm). As
for the blue star, number 8, in the background of NGC 2660 it may
be a member of the Local Arm though it is still possible for it to
become one of the innermost members of the Perseus Arm. Fig. 15
shows the location of another remote blue star, number 10, seen in
the background of Ruprecht 47 at 9.5 kpc from the Sun. We have
already commented above that this star may become member of the
innermost part of the Outer Arm. Nevertheless, what makes this
object a singular star is its location along the Galactic plane.
In fact, at $l= 240^\circ-250^\circ$ the warp is quite pronounced
(V\'azquez et al. 2008) but some young open clusters as Haffner 18
and Haffner 19 and blue stars in the background of Haffner 21 and
Trumpler 9 can be found (see Carraro et al. 2015). Gathering all
this information we interpret that the presence of this very
remote young star in the Galactic plane is not an isolated fact
but it is a strong indication that the Galactic thin disk is not
only warped, but also flared, like the thick disk. Quite recently,
Xu et al (2013) have presented evidences of a large number of star
formation regions with trigonometric parallaxes that extend the
Local Arm back to the First Galactic Quadrant. Xu et al. (2013)
stated also a series of questions (not answered yet) concerning
the interpretation of the Local Arm: 1.- it is a branch of the
Perseus Arm, 2.- it is part of a major arm or 3.- it is an
independent spiral arm segment. Unfortunately, we cannot speculate
about the nature of the Local Arm because our observations were
designed just to establish the existence and the spatial
distribution of young objects in the TGQ. More observations
(photometry and spectroscopy) for a precise estimation of cluster
parameters are needed in order to reach a plausible answer to
these questions from an optical point of view. Meanwhile, a
comparison of our Fig. 15 with their Fig. 10 reveals that our
findings are in good agreement with theirs in terms of the
distribution of young population throughout the Local Arm in TGQ.

\section{Conclusions}

We presented in this article the results of a photometric and
spectroscopic survey carried out in five cluster located at
$240^\circ < l < 275^\circ$ and $0^\circ > b > -6^\circ$. This
region comprises mostly the Vela Gum region and includes part of
the Puppis association.

One of our clusters, Ruprecht 60, has been studied with $UBV$
photometry for the first time and its parameters are now improved.
Other of the clusters, Ruprecht 20, is not a cluster since neither
from star counts used to construct its radial density profile nor
from the photometric diagrams we can conclude that it is a true
entity. It is just a random stellar fluctuation.

For the other three clusters, Ruprecht 47, NGC 2660 and NGC 2910
we have improved also their parameters.

From the point of view of identification of large stellar
structures in this portion of the Galaxy we have found evidences
that favor the presence in Puppis, at $b^\circ =-4$, of a sort of
bridge connecting two regions defined by Kaltcheva \& Hilditch
(2000). Ruprecht 47 is definitely a member of Puppis OB2 and in
its field we also found two blue stars related with
Puppis OB1. NGC 2910 becomes member of Vel OB1 and also a
couple of blue stars with no relation to NGC 2910.

Ruprecht 60 and NGC 2660 are too old objects with no genetic
relation to the associations they are seen against.

As for two peculiar isolated stars we report the presence of a
very distant star seen against the field of Ruprecht 47 placed at
almost 10 kpc from the Sun. Ruprecht 47 is along the formal
Galactic plane ($b^\circ =-0.188$) and this remote star in
the background may belong to the innermost part of the
Outer Arm. Another interesting star has been found in the field of
NGC 2660 at near 5.7 kpc from the Sun. This star could be member
of the innermost part of the Perseus Arm or a far member of the Local Arm. 
As for the grand design structures in
the TGQ we are in the position to confirm the angular extension of
the Local Arm from $220^\circ$ to $275^\circ$. Our findings are in
good agreement with previous studies (see Carraro et al. 2015) we
have already performed in the sense that our evidences indicate
that the Galactic plane is not only warped but also flared. This
is, thin disk population shows a pattern similar to the shown by
the thick disk. Moreover, there is a good spatial agreement with
Xu et al. (2013) findings. Finally, the kind of observations we
performed do not allow to speculate about the nature of the Local
Arm but our evidences -collected in several papers as indicated in
Section 1 and concerning strictly the TGQ- suggest that the Local
Arm may even cross the Perseus Arm and reach the Outer Arm.

Acknowledgements: We warmly acknowledge the CASLEO staff for the
very professional support along the years. One of us, RAV, wishes
to thank to Dr. Garrison for the allocation of observing time at
the University of Toronto Southern Observatory in Las Campanas
Observatory, Chile. EG, GS and RAV thank the financial support
from CONICET PIPs 1359 and 5970 and from La Plata Observatory. We
especially acknowledge to our anonymous referee for the careful
and collaborative report that greatly contributed to improve the
value of our manuscript. This publication makes use of the Two
Micron All Sky Survey, which is a joint project of the University
of Massachusetts and the Infrared Processing and Analysis
Center/California Institute of Technology, funded by the National
Aeronautics and Space Administration and the National Science
Foundation. The Digitized Sky Survey was produced at the Space
Telescope Science Institute under U.S. Government grant NAG
W-2166. The images of these surveys are based on photographic data
obtained using the Oschin Schmidt Telescope on Palomar Mountain
and the UK Schmidt Telescope. The plates were processed into the
present compressed digital form with the permission of these
institutions.



%% References


% \bibliographystyle{elsarticle-num}
% \bibliography{<your-bib-database>}





\end{document}